\documentclass[12pt]{article}
\usepackage{amsfonts}
\usepackage[sort,compress]{natbib}
\usepackage{framed}
\usepackage{subcaption}
\usepackage{hyperref}
\usepackage{amsmath,amssymb,amsfonts,bm}
\usepackage{amsthm,amsfonts}
\usepackage{mathrsfs}
\usepackage{xcolor}
\usepackage{graphicx}
\usepackage{comment}

\usepackage[margin=1in]{geometry}

\usepackage[linesnumbered,ruled]{algorithm2e}

\usepackage{empheq}

\usepackage{mathtools}

\mathtoolsset{showonlyrefs}

\usepackage{xr}
\newcommand{\llll}{\boldsymbol \lambda}
\newcommand{\ttt}{\boldsymbol \theta}
\newcommand{\FF}{\mathcal{F}}

\newtheorem{assump}{}

\numberwithin{equation}{section}

\newtheorem{theorem}{Theorem}
\newtheorem{lemma}[theorem]{Lemma}

\newtheorem{coro}[theorem]{Corollary}

\theoremstyle{definition}

\theoremstyle{remark}

\DeclareMathOperator{\argmin}{argmin}

\newcommand{\RR}{\mathbb{R}}

 \newcommand{\PP}{\mathbb{P}}
 \newcommand{\QQ}{\mathbb{Q}}
\newcommand{\EE}{\mathbb{E}}
\newcommand{\E}{\mathbb{E}}

\newcommand{\norm}[1]{\lVert#1\rVert}

\renewcommand{\argmin}{\mathop{\mathrm{arg\,min}{}}}
\newcommand{\argmax}{\mathop{\mathrm{arg\,max}{}}}

\definecolor{DSgray}{cmyk}{0,1,0,0}
\newcommand{\Authornote}[2]{{\small\textcolor{DSgray}{\sf$<${  #1: #2
}$>$}}}

\newcommand{\ynote}[1]{{\Authornote{YC}{#1}}}

\newcommand{\tup}[0]{{{3 }}}

\let\hat\widehat
\let\tilde\widetilde

\author{    Xi Chen \\ 
    Stern School of Business, 
    New York University \\
  Yunxiao Chen \hspace{.2cm}\\
Department of Psychology, Emory University\\
Xiaoou Li\\
  School of Statistics, University of Minnesota\\
    }
  \date{\today}

\begin{document}
\title{Asymptotically Optimal Sequential Design for Rank Aggregation}
\maketitle

\begin{abstract}

A sequential design problem for rank aggregation is commonly encountered in psychology, politics, marketing, sports, etc.
In this problem,
a decision maker is responsible for ranking $K$ items
by sequentially collecting pairwise noisy comparison from judges.
The decision maker needs to choose a pair of items for comparison in
each step, decide when to stop data collection, and make a final decision after stopping, based on a sequential flow of information.
Due to the complex ranking structure, existing sequential analysis methods are not suitable.

In this paper, we formulate the problem under a Bayesian decision framework and propose sequential procedures that are
asymptotically optimal. These procedures achieve asymptotic optimality by
seeking for a balance between exploration (i.e. finding the most indistinguishable pair of items) and exploitation (i.e. comparing the most indistinguishable pair based on the current information).
New analytical tools are developed for proving the asymptotic results, combining advanced change of measure techniques for handling the level crossing of likelihood ratios and classic large deviation results for martingales, which are of separate theoretical interest in solving complex sequential design problems. A mirror-descent algorithm is developed for the computation of the proposed sequential procedures.

\end{abstract}




\section{Introduction}

This paper considers a sequential design problem for rank aggregation.
In this problem, a decision maker is responsible for ranking $K$ items
by adaptively collecting noisy outcome of pairwise comparison from judges.
The decision maker needs to choose a pair of items for comparison in
each step, decide when to stop data collection, and make a final decision after stopping, based on a sequential flow of information.
Due  to its special structure, this problem cannot be formulated and solved by
existing sequential adaptive design methods \citep{chernoff1959sequential,Naghshvar:13}.

Sequential rank aggregation has a wide range of applications, including social choice \citep{saaty2012possibility}, sports \citep{elo1978rating}, search rankings \citep{page1999pagerank}, etc. Pairwise comparison is the most popular approach for rank aggregation, as sufficient evidence from cognitive psychology suggests that people make more accurate judgement
when making pairwise comparisons (i.e., given a pair of items and asked to indicate which item is preferred to the other) as compared to multi-wise comparison \citep{blumenthal1977process}
and some applications such as chess gaming have a natural form of pairwise comparison.

It is intuitive that in this sequential design problem, one should choose the  most indistinguishable pair of items to compare based on the current information and stops when the ambiguity of all item pairs falls below a certain level.
The focus of this paper is to make this intuition rigorous
by formulating the problem under a Bayesian decision framework and
show that this intuition leads to sequential design procedures that are asymptotically optimal, where the
 notion of asymptotic optimality follows \cite{chernoff1959sequential} that is
widely used in sequential analysis \citep{Siegmund85Seq,Lai01,Tartakovsky14Seq,Schwarz1962}.
In our formulation, each item $k$ is represented by a parameter $\theta_k$, which determines its underlying true rank among $K$ items.  For example, the parameter $\theta_k$ can be viewed as the quality score for item $k$, and item $i$ has a higher rank than item $j$ if and only if $\theta_i > \theta_j$. The pairwise comparison of items $i$ and $j$ follows a probabilistic comparison model (e.g., \cite{Thurstone:27,Bradley52,Luce59}) parameterized by $\theta_i$ and $\theta_j$.
A sequential procedure chooses a pair $(i,j)$ for the next comparison in each stage and decides the stopping time $T$. Upon stopping,  the final decision is to choose the global rank $R:=(R_1, \ldots, R_K)$ from the set of all permutations of $\{1, 2, ..., K\}$.
The loss function of this sequential design problem is defined by combining the
cost of data collection and
the Kendall's tau distance  \citep{kendall1948rank} between the decision $R$ and the underlying scores $(\theta_1, ..., \theta_K)$:
\begin{equation}\label{eq:loss}
  \sum_{i < j} I({\theta_i>\theta_j)}I(R_i > R_j) + I(\theta_i<\theta_j)I(R_i < R_j)+ c T,
\end{equation}
where the constant $c>0$ indicates the  relative cost of each comparison and $I(\cdot)$ denotes an indicator function.


Although according to the final decision, our problem seems to be a multi-hypotheses sequential testing problem with adaptive experiment selection considered in \cite{Naghshvar:13},  there exist fundamental differences. First, \cite{Naghshvar:13} only consider
simple hypotheses, while the ranking problem, when viewed as a
multi-hypothesis testing problem, consists of composite hypotheses.
Second, typically $0-1$ loss is considered for measuring the decision accuracy in multi-hypothesis testing, while our problem has a more complex loss function based on the Kendall's tau distance that is tailored to rank aggregation. Our problem is also a substantial generalization of classical sequential test of two composite hypotheses \citep{Schwarz1962,Kiefer1963,lai1988nearly}. In particular, when the
number of items is two ($K=2$), our problem degenerates to testing two composite hypotheses without adaptive experiment selection.

\subsection{Main contribution} In this paper, we develop new sequential analysis methods to conduct sequential experiments for pairwise comparisons and to balance the ranking accuracy and cost. The main methodology and theoretical contributions  of the paper are summarized as follows,
\begin{itemize}
\item Under a Bayesian decision framework and under a large class of parametric pairwise comparison models,
we derive an asymptotic lower bound (Theorem~\ref{thm:lb}) for the Bayes risk of all possible sequential ranking policies. Note that
the Bayes risk of the sequential rank aggregation problem, which combines
the expected Kendall's tau distance and the expected sample size,  is more complex than that of traditional sequential hypothesis testing problems.
\item  We propose two sequential ranking policies. In particular, we provide two choices of stopping rule and a class of randomized pair selection rules. We quantify the expected Kendall's tau and the sample size of the proposed methods (Theorems \ref{thm:error} and \ref{thm:stopping-time}) and show that the Bayes risks match the asymptotic lower bound, which further implies that the proposed methods are asymptotically optimal (Corollary \ref{cor:opt}).
Our randomized pair selection rule utilizes an epsilon-greedy strategy to balance the exploitation (i.e., choosing the best pair for comparison based on current information) and exploration (i.e., randomly selecting pairs to gain information about underlying parameters $\{\theta_k\}_{k=1}^K$). The exploration is critical for learning the rank, while the exploitation is critical for saving the comparison cost. 
\begin{itemize}
	\item For the exploration, we quantify the impact of the exploration rate on the estimation of model parameters and provide an exponential probability bound as an auxiliary result (Lemma~\ref{lemma:mle}).
	\item For the  exploitation, we consider a randomized adaptive selection rule (see Section \ref{sec:method}). Specifically, in each step, the probability of selecting each pair is obtained by solving
a saddle point optimization problem.
We further develop a mirror descent algorithm for solving the optimization (see Section \ref{sec:comp}).
\end{itemize}
\item Technically, we develop new analytical tools for quantifying the level crossing probability of a random function (e.g. likelihood function, martingale, or sub-martingale) double-indexed by model parameters and the  sample size.  As such a probability tends to zero, the problem falls into the rare-event analysis domain, where an exact exponential decay rate is challenging to obtain. Traditional methods, such as the ones adopted in \cite{Naghshvar:13,chernoff1959sequential}, are based on exponential change-of-measure of the log-likelihood ratio statistics, and are not directly applicable to the ranking problem considered here. The method we use in the proof combines a mixture-type of change-of-measure method  recently proposed in \cite{adler2012efficient,li2016chernoff,li2015rare} and large deviation results for martingales.

\end{itemize}


\subsection{Related works} Sequential hypothesis testing, initiated by the seminal works of  \cite{Wald45} and \cite{wald1948optimum},  is an important area of statistics for processing data taken in a sequential experiment, where the total number of observations is not fixed in advance.  A sequential test is characterized by  two components: (1) a stopping rule that decides when to stop the data collection process, and (2) a decision rule on choosing the hypothesis upon stopping.
A large body of literature on sequential tests with two hypotheses has been developed, a partial list of which includes \citep{Schwarz1962,Kiefer1963,Hoeffding1960,lai1988nearly}.
Sequential testing with more than two hypotheses and sequential  multiple testing have been extensively studied in recent decades (see, e.g., \cite{draglia1999multihypothesis,dragalin2000multihypothesis,mei2010efficient,Xie:13,Song:17:mt}).
For a comprehensive review on sequential analysis, we refer the readers to the surveys and books
\citep{Siegmund85Seq,Lai01,Hsiung:04,Tartakovsky14Seq} and references therein.
 In addition to optimizing over the stopping rule and final decision, \cite{chernoff1959sequential} first introduces the adaptive design into the sequential testing framework, followed by a large body of literature, see, e.g.  \cite{albert1961sequential,tsitovich1985sequential,Naghshvar:13,Nitinawarat:15}.
Sequential analysis finds many applications in different disciplines, including clinical trials, educational testing, and industrial quality control (see, e.g., \cite{bartroff2008modern,bartroff2008efficient,bartroff2013sequential,lai2004power,wang2016hybrid,ye2016sequential}).

The rank aggregation problem has been an active research problem in recent years (see, e.g., \cite{Negahban12RankCentrality,Hajek14,Shah15Sto} and references therein), which finds many applications to social choice, tournament play, search rankings, advertisement placement, etc. With the advent of crowdsourcing services, one can easily ask crowd workers to conduct comparisons among a few objects in an online fashion at a low cost \citep{Chen:13,Chen:16:Ranking}. Although rank aggregation has been extensively studied in the machine learning community, it has not been investigated under the sequential analysis framework. The techniques developed in this work will enable a sequential rank procedure with optimal stopping and adaptive design.

Our problem is also related to, but substantially different from,  the  selecting and ranking problem \citep{gupta1965some,bechhofer1968sequential,gupta2002multiple}, which collects data from
$K$ populations and studies the sequential design for finding the population with the largest mean. Due to the different objectives,  the techniques used for selecting and ranking, such as sequential elimination, are not applicable to our problem.

\subsection{Paper Organization} The rest of the paper is organized as follows. In Section \ref{sec:setup}, we introduce the setup of the problem.
Section \ref{sec:method} presents the proposed policies and the  theoretical results, and provides further discussions.
Section~\ref{sec:simu}
presents the simulation results, followed by concluding remarks in Section \ref{sec:conclusion}.
Technical proofs for the Theorems are provided in the Section~\ref{sec:proof}. Proofs for all the lemmas are provided in the supplementary material. 

\section{Problem Setup} \label{sec:setup}

We first introduce the comparison model and formulate the sequential ranking problem. Consider the task of inferring a global ranking over $K$ items. Let  $\mathcal{A}= \{(i,j): i,j\in\{1,...,K\}, i< j \}$ be the set of pairs for comparison. At each time $n$ ($n=1,2, \ldots$), a pair $a_n:=(a_{n,1}, a_{n,2}) \in \mathcal{A}$ is selected for comparison.
For example, $a_2=(1,2)$ means that items 1 and 2 are compared at time two.
The comparison outcome is denoted by a random variable $X_n \in \{0, 1\}$, where $X_n=1$ means item $a_{n,1}$ is preferred to item $a_{n,2}$ and $X_n=0$ otherwise. The comparison outcome $X_n$ is assumed to follow a ranking model, such as the widely used Bradley-Terry-Luce (BTL) model \citep{Bradley52,Luce59}
and Thurstone model \citep{Thurstone:27}. Such a ranking model assumes that
each item is associated with an unknown latent score $\theta_i \in \mathbb{R}$, for $i=1,\ldots, K$, where
the global rank of the $K$ items is given by the rank of $\theta_1$, ...., $\theta_K$.
The distribution of $X_n$ is determined by $\theta_i$ and $\theta_j$, when comparing pair $(i, j)$. For example,  given pair $a_n:=(a_{n,1}, a_{n,2})$, the BTL model assumes that,
\begin{equation}\label{eq:BTL}
\begin{aligned}
& \PP(X_n=1) = \frac{\exp(\theta_{a_{n,1}})}{\exp(\theta_{a_{n,1}}) +\exp(\theta_{a_{n,2}}) };\\
& \PP(X_n=0) = \frac{\exp(\theta_{a_{n,2}})}{\exp(\theta_{a_{n,1}}) +\exp(\theta_{a_{n,2}}) }.
 \end{aligned}
\end{equation}
Under this model, $\theta_{a_{n,1}} > \theta_{a_{n,2}}$ means that item $a_{n,1}$ is preferred to item $a_{n,2}$, reflected by $\PP(X_n=1)  > 0.5$. A common feature for many comparison models is that the distribution of the comparison of items  $i$ and $j$
only depends on the pairwise differences $\theta_{i}-\theta_{j}$.
Consequently,
such models are not identifiable up to a location shift.
To overcome this issue, we fix $\theta_1=0$ and treat $\ttt=(\theta_2,...,\theta_{K})$ as the unknown model parameters.
The result of this paper applies to a wide class of comparison models and thus we denote the probability mass function of the comparison outcome $x$ given pair $a$ as $f_{\ttt}^a(x)$.


We now describe components in a {sequential design for rank aggregation}: an adaptive selection rule $A$, a stopping time $T$, and a decision rule $R$ on the global rank.
For the adaptive selection rule $A$, we
consider the class of randomized adaptive selection rules, which contains deterministic selection rules as special cases.
In particular, let $A=\{\llll_n: n=1,2,... \}$, where $\llll_n=(\lambda_n^{i,j})_{(i,j)\in \mathcal{A}} \in \Delta$ denotes the probability of selecting the pair $(i,j)$.
Here, $\Delta=\{(\lambda^{i,j}: \sum_{(i,j)\in \mathcal{A}} \lambda^{i,j} =1, \; \lambda^{i,j} \geq 0 \}$ is a probability simplex over $K(K-1)/2$ pairs.
At each time $n$, a pair
$a_n$ is selected according to the categorical distribution $\llll_n$,
where
$\llll_n$ adapts to the  filtration sigma-algebra generated by the selected pairs and the observed outcomes, that is, $\mathcal{F}_{n}=\sigma(X_1,...,X_{n-1}, a_1,...,a_{n-1})$.
The adaptive comparison process will stop at time $T$,
a stopping time with respect to the filtration $\{\mathcal{F}_{n}\}_{n \geq 0}$.
It is worthwhile to note that the random stopping time $T$ is also
the number of samples being collected.
Upon stopping, one needs to make a decision $R:=(R_1, \ldots, R_K)$, the global ranking of the $K$ items.
For example, when $K=3$, $R=(3,1,2)$ means that one decides $\theta_{2}>\theta_3>\theta_1$.
We further denote $P_K$ the set of permutations over $\{1,\ldots K\}$
and thus $R \in P_K$.
The adaptive selection rule  $A=\{\llll_n: n=1,2,... \}$, the stopping time $T$, and the decision $R$ together form a \emph{sequential ranking policy}, denoted by $\pi=(A, T,R)$.

The performance of a sequential ranking policy is measured via its ranking accuracy and the expected stopping time.
Specifically, we measure the ranking accuracy by Kendall's tau distance \citep{kendall1948rank}, which is one of the most widely used measures for ranking consistency.
More precisely,  for each $R=(R_1,...,R_K)\in P_K$, we convert it to the binary decisions over pairs $\{R_{i,j}\in\{0,1\}: i,j\in \{1,...,K\}, i <j \}$, where $R_{i,j}=I(R_i<R_j)$, and $R_{i,j}=1$ means that item $i$ is preferred to item $j$. For example,  if $R=(3,1,2)$, we have  $R_{1,2}=0$ and $R_{2,3}=1$. The Kendall's tau distance between $R$ and the true ranking induced by $(\theta_1,\ldots, \theta_K)$ is defined by
\begin{equation}\label{eq:loss-k}
	L_{K}(\{R_{i,j}\})=\sum_{i<j} I({\theta_i>\theta_j)}(1-R_{i,j}) + I(\theta_i<\theta_j)R_{i,j}.
\end{equation}
On the other hand, the loss function associated with the random sample size $T$ is defined as,
\begin{equation}\label{eq:cost}
	L_{c}(T)=c\times T,
\end{equation}
where the constant $c>0$
indicates the \emph{relative} cost of conducting one more pairwise comparison. The choice of $c$ depends on the nature of the ranking problem. Generally, if obtaining each sample is expensive comparing to the cost due to the inaccuracy of the ranking, then a large $c$ will be chosen and vise versa.
Note that $c$ is not a tuning parameter to optimize over.




We define the risk associated with a sequential ranking policy under the Bayesian decision framework,
in which the model parameter $\ttt$ is assumed to be random and follows a prior distribution.
To avoid confusion, we write $\Theta$ when $\ttt$ is viewed as random,
and denote by $\rho(\ttt)$ the prior density function of $\Theta=(\Theta_2,...,\Theta_K)$.  Recall that we have fixed $\Theta_1=0$ to ensure identifiability.
The Bayes risk combines the risks associated with Kendall's tau distance of the decision and the sampling cost,
\begin{align} \label{eq:risk}
       V_c(\rho,\pi) &=\E^{\pi} \left(L_K(\{R_{i,j}\})+L_c(T) \right)\\
                  &= \E^{\pi}\Big\{\sum_{i<j} I({\Theta_i>\Theta_j)}(1-R_{i,j}) + I(\Theta_i<\Theta_j)R_{i,j} \Big\}
      +c \E^{\pi} T, \nonumber
\end{align}
where the expectation $\E^{\pi}$ is taken under the policy $\pi$, with respect to the randomness of the selected pairs, the observed comparison results, and the stopping time. Of particular interest is the minimum risk under the optimal sequential ranking policy given the prior distribution of $\Theta$ and sampling cost $c$
\begin{equation}\label{eq:V*}
	V_c^{*}(\rho)=\inf_{
\pi} V_c(\rho,\pi).
\end{equation}

For any given cost $c$,
obtaining an analytical form of an optimal policy that achieves $V^{*}(\rho, c)$
is typically infeasible. Following the literature of sequential analysis, a policy is usually evaluated by
the notion of
\emph{asymptotic optimality} \citep{chernoff1959sequential}.
In particular, a policy $\pi$ is said to be asymptotically optimal if
\begin{equation}\label{eq:asymptotic-optimal}
	\lim_{c\to 0}\frac{V_c(\rho,\pi)}{V_c^{*}(\rho)}=1,
\end{equation}
i.e. when the relative sampling cost converges to 0.

%
%
%
%
%
%
%
%
%





\section{Sequential Policies and Asymptotic Optimality}\label{sec:method} In Section~\ref{subsec:policy}, we propose two sequential ranking  policies $\pi_1$ and $\pi_2$. Then the asymptotic optimality of the two policies is presented in Section~\ref{subsec:optimal}.
Further discussions are provided in Section~\ref{subsec:discusion}.

\subsection{Two Sequential Policies}\label{subsec:policy}

We first introduce some notations. Let $W$ be the support of the prior probability density function $\rho$, i.e., $W=\overline{\{ \ttt: \rho(\ttt)>0\}}$, where $\bar{E}$ denotes the closure of a set $E$.
We further define the set $W_{i,j}=\{ \ttt: \theta_i\geq\theta_j  \}\cap W$ for all $i,j\in \{1,...,K\}$. It is worthwhile to note that $W_{i,j}$ and $W_{j,i}$ are different sets and their union is the set $W$.
Given a sequence of selected pairs $a_1,...,a_n$ and observed comparisons $X_1,...,X_n$, the log-likelihood function is  defined as,
\begin{equation}\label{eq:likelihood}
	l_n(\ttt)=\sum_{i=1}^n \log f^{a_i}_{\ttt}(X_i),
\end{equation}
and the corresponding maximum likelihood estimator $\hat{\ttt}^{(n)}=(\hat{\theta}^{(n)}_2,...,\hat{\theta}^{(n)}_K)$
is
\begin{equation}\label{eq:MLE}
  \hat{\ttt}^{(n)}=\arg\sup_{\ttt \in W} l_{n}(\ttt).
\end{equation}

We then introduce two stopping times based on the generalized likelihood ratio statistic,
\begin{equation}\label{eq:t-1}
	T_1=\inf\Big\{n > 1 : \sum_{(i,j) \in \mathcal{A}}
	    \exp\{-|\sup_{{\ttt}\in W_{i,j}}l_n({\ttt})-\sup_{\ttt\in
        W_{j,i}}l_n(\ttt) | \}\leq e^{-h(c)} \Big\},
\end{equation}
and
\begin{equation}\label{eq:t-2}
	T_2=\inf\Big\{n > 1: \min_{(i,j) \in \mathcal{A}}|\sup_{\ttt\in W_{i,j}}
        l_n(\ttt)-\sup_{\ttt\in W_{j,i}}
        l_n(\ttt)|
    	\geq h(c)\Big\},
\end{equation}
where $h(c)= |\log c|(1+|\log c|^{-\alpha})$ for some constant $\alpha\in(0,1)$ and $c$ is  the relative cost introduced in \eqref{eq:cost}. We note that $T_2$ is obtained by replacing the summation in $T_1$ by maximization and taking log and minus on both sides. Upon stopping, the  decision about the global rank is made according to the rank of MLE at the stopping time $T$ ($T=T_1$ or $T_2$). That is,
\begin{equation}\label{eq:decision}
	R = r(\hat{\ttt}^{(T)}),
\end{equation}
where the function $r(\ttt): \mathbb{R}^{K-1} \to P_K$ gives the rank of { $(0, \theta_2,...,\theta_K)$.}
More precisely,
$r(\ttt)=(r_1,\ldots, r_K) \in P_K$, satisfying
 $\theta_{r_1} \geq \theta_{r_2} \geq \ldots \geq \theta_{r_K}$, where $\theta_1 = 0$.
We provide more intuitions behind the stopping rules $T_1$ and $T_2$ and the decision rule $R$ in Section~\ref{sec:connections}.

We proceed to the randomized selection rule $A$,
which is obtained by solving an optimization program.
For a given $\ttt$, we define function $D(\ttt)$,
\begin{equation}\label{eq:obj}
	D(\ttt)=\max_{\llll\in\Delta} \min_{\tilde{\ttt}\in W: r(\tilde{\ttt})\neq r(\ttt)} \sum_{(i,j)}\lambda^{i,j}D^{i,j}(\ttt\|\tilde{\ttt}),
\end{equation}
where $D^{i,j}(\ttt\|\tilde{\ttt})$ is the Kullback-Leibler (KL) divergence from $f^{i,j}_{\tilde{\ttt}}(\cdot)$ to $f^{i,j}_{\ttt}(\cdot)$, i.e.
\[
D^{i,j}(\ttt\|\tilde{\ttt}):=\sum_{x \in \{0, 1\}} f^{i,j}_{\ttt}(x) \log\frac{f^{i,j}_{\ttt}(x)}{f^{i,j}_{\tilde{\ttt}}(x)}.
\]
We further define
\begin{equation}\label{eq:saddle_obj}
  \llll^*(\ttt)=\argmax_{\llll\in\Delta} \min_{\tilde{\ttt}\in W: r(\tilde{\ttt})\neq r(\ttt)} \sum_{(i,j)}\lambda^{i,j}D^{i,j}(\ttt\|\tilde{\ttt}),
\end{equation}
and
\begin{equation}\label{eq:policy}
  \hat{\llll}_{n}= (\hat \lambda_n^{i,j}) = \llll^*(\hat{\ttt}^{(n-1)}).
\end{equation}
That is, $\llll^*(\ttt)$ is the solution to the optimization problem \eqref{eq:obj}, and $\hat{\llll}_{n}$ is the solution to the optimization problem given the MLE based on the previous $n-1$ samples. 
The objective function in \eqref{eq:obj} is a weighted KL divergence for all pairs with the weights $\lambda^{i,j}$. The inner minimization problem is taken over all the parameter vector $\tilde{\ttt} \in W$, for which the
induced rank $r(\tilde{\ttt})$ is different from that of $\ttt$.
At each time $n$, given the MLE $\hat{\ttt}^{(n-1)} $,
we compute $\hat{\llll}_{n}$, which is the maximizer of $\llll \in \Delta$ in $D(\hat{\ttt}^{(n-1)})$.
We elaborate on the intuition behind the optimization in \eqref{eq:obj}.
First, for each $\ttt$, $\sum_{(i,j)}\lambda^{i,j}D^{i,j}(\ttt\|\tilde{\ttt})$
gives the drift of the log-likelihood ratio statistics between $f_{\ttt}$ and $f_{\tilde{\ttt}}$ under the model $f_{\ttt}$ and a randomized sampling scheme specified by $\llll$, which is also
the mutual information between $f_{\ttt}$ and $f_{\tilde{\ttt}}$ when the pair is selected according to $\llll$.   Minimizing the inner part with respect to $\tilde{\ttt}$ over the set $\{\tilde{\ttt}\in W:r(\tilde{\ttt})\neq r(\ttt)\}$ provides a measure on the distinguishability of the rank of $\ttt$ under the sampling scheme  $\llll$. Second, if the true model parameter is $\ttt$, we would like to choose a sampling scheme $\llll$ such that it provides the highest distinguishability obtained by the first step. Thus, we perform maximization in the outer part of \eqref{eq:obj}. Finally, as the true model parameter $\ttt$ is unknown, we will replace $\ttt$  by the MLE based on the current information.
In Section \ref{sec:comp}, we provide a mirror descent algorithm for solving \eqref{eq:obj}.

Unfortunately, directly using
$\hat{\llll}_n$ in the selection rule $A$ as
the choice probability does not guarantee the asymptotic optimality. This is because
$\hat{\llll}_n$ does not guarantee sufficient exploration of all item pairs, which may lead to the imbalance between the exploration and exploitation for the sequential procedure.
To fix this issue, we combine
$\hat{\llll}_n$  with an $\epsilon$-greedy approach which is widely used in balancing exploration and exploitation in multi-armed bandit and decision-making problems (see, e.g., \cite{Watkins:89}).
Specifically, an exploration probability $p \in (0,1)$ is chosen, which is typically small and
may be chosen depending on the value of the relative sampling cost $c$.
At each time $n$, with probability $p$, we select the next pair uniformly from $\mathcal{A}$. With probability $1-p$, the next pair is selected according to the categorical distribution specified by $\hat{\llll}_{n}$. In other words, for each pair $(i,j)$, the choice probability of the selection rule at time $n$ is given by
$$\lambda_{n}^{i,j}= p \frac{2}{K(K-1)} +(1-p) \hat{\lambda}_{n}^{i,j}.$$

We call the above selection rule $A_p$, where the subscript emphasizes its dependence on the exploration rate $p$.
The two proposed sequential ranking policies are defined by $\pi_1 := (A_p, T_1, R)$ and $\pi_2 := (A_p, T_2, R)$.
The proposed sequential ranking policies are summarized in Algorithm \ref{algo:seq} and the computation for solving
\eqref{eq:saddle_obj} will be discussed in Section~\ref{sec:comp}.
The proofs of the theoretical results are provided in Section \ref{sec:proof}.

\begin{algorithm}[!ht]
\begin{flushleft}
\caption{Sequential Ranking Policy}
\textbf{Input}: The probability mass (density) function $f_{\ttt}^a(x)$ for any pair $a \in \mathcal{A}$,
the probability $p \in (0,1)$ in  $\epsilon$-greedy, and the support $W$ of $\rho(\ttt)$.

\textbf{Initialization}: Uniformly sample a pair $a_1$ at random and observe the comparison outcome $X_1$.

\textbf{Iterate} For $n=2,3,\ldots$ until the stopping time $T$ in \eqref{eq:t-1} (or \eqref{eq:t-2}) is  reached.
 \begin{enumerate}
   \item Compute the MLE based on the previous $n-1$ comparisons:
   \[
      \hat{\ttt}^{(n-1)}=\arg\sup_{\ttt \in W} l_{n-1}(\ttt).
   \]

   \item Compute
   \begin{equation}\label{eq:choice_prob}
      \hat{\llll}_n = \argmax_{\llll \in \Delta} \min_{\tilde{\ttt}\in W: r(\tilde{\ttt})\neq r(\hat{\ttt}^{(n-1)})} \sum_{(i,j) \in \mathcal{A}}\lambda^{i,j}D^{i,j}(\hat{\ttt}^{(n-1)} \|\tilde{\ttt}).
   \end{equation}

   \item Flip a coin with head probability $p$.
   \begin{itemize}
         \item If the outcome is  head, select the pair $a_n$ uniformly at random over all pairs from $\mathcal{A}$.

         \item Otherwise, select the pair $a_n$ according to the categorical distribution specified by $\hat{\llll}_n$.
   \end{itemize}

   \item Observe the comparison result $X_n$ and update the likelihood function $l_n(\ttt)$.
 \end{enumerate}
\textbf{Output}: The rank $R=r(\hat{\ttt}^{(T)})$, i.e., the global rank induced by $\hat{\ttt}^{(T)}$.
 \label{algo:seq}
\end{flushleft}
\end{algorithm}


\subsection{Asymptotic Optimality}\label{subsec:optimal}
This section contains the main results of the paper, including (1)
a lower bound on the risk of a general sequential ranking procedure, and
(2)  
theoretical analysis on the proposed procedures, which leads to their asymptotic optimality.
As a by-product, an exponential deviation bound for the  MLE over a moving window is also presented. The assumptions for our results are described and discussed.

\paragraph{Notations} Throughout the rest of the paper, we  write $a_c=O(b_c)$ for two sequences $a_c$ and $b_c$ if $|a_c|/|b_c|$ is bounded, uniformly in $\ttt$, as $c\to 0$. Similarly, we write $a_c=\Omega(b_c)$ if $a_c>0$, $b_c>0$ and $b_c=O(a_c)$. We will also write $a_c=o(b_c)$ if $a_c/b_c\to 0$ uniformly in $\ttt$.

\paragraph{Main results}


Let us first describe our assumptions. 
For technical needs, we make some regularity conditions on the prior distribution $\rho(\ttt)$. {Recall that we have fixed $\theta_1=0$ and let $\ttt=(\theta_2,...,\theta_K)\in\RR^{K-1}$ be the unknown model parameters.}
\begin{assump}\label{assump:compact}
	The support {$W:=\overline{\{\ttt\in \mathbb{R}^{K-1}: \rho(\ttt)>0 \} }$ is a compact set in $\mathbb{R}^{K-1}$}, where $\bar{E}$ denotes the closure of a set $E$.  In addition, for {any permutation $\sigma \in P_{K}$},
	$(\{\ttt \in \mathbb{R}^{K-1}:
	r(\ttt)=\sigma
	\}\cap W)^\circ\neq\emptyset$,
  where $E^{\circ}$ denotes the interior of a set $E$.
\end{assump}
\begin{assump}\label{assump:reg-w}
	 There exists a constant $\delta_b>0$ such that for all $s>0$,
	{$
	m(B(\ttt,s)\cap W)\geq \min\{\delta_b s^{K-1}, 1\},
	$}
	where $B(\ttt,s)$ denotes the open ball centered at $\ttt$ with radius $s$ and $m(\cdot)$ denotes the Lebesgue measure. 
\end{assump}
\begin{assump}\label{assump:f}
	The function $\log f_{\ttt}^{a}(x)$ is continuously differentiable  in $\ttt$ for all $x$ uniformly. That is,
  \begin{equation}
    \sup_{\ttt\in W, a\in\mathcal{A},x} \|\nabla_{\ttt} \log f_{\ttt}^{a}(x)\|<\infty.
  \end{equation}
\end{assump}
\begin{assump}\label{assump:separate}
$
\min_{
	\ttt,\tilde{\ttt}\in W: r(\tilde{\ttt})\neq r(\ttt)
	}	\max_{(i,j)}D^{i,j}(\ttt\|\tilde{\ttt})>0.
$
\end{assump}
\begin{assump}\label{assump:prior}
	$\inf_{\ttt\in W^{\circ}} \rho(\ttt)>0$ and $\sup_{\ttt\in W} \rho(\ttt)<\infty$.
\end{assump}
We provide some remarks on the above
regularity assumptions.
Assumption~\ref{assump:compact} requires  the prior distribution for $\Theta$ to have a bounded support, which has a non-empty interior for each rank. Assumption~\ref{assump:reg-w} avoids the support $W$ being singular.
Assumption~\ref{assump:f} requires the smoothness of the likelihood function. Assumption~\ref{assump:separate}  requires that there is no tie in the support of the prior distribution.
This is a standard assumption in sequential analysis, which corresponds to the classic ``indifference zone'' assumption in sequential hypothesis testing
\citep{Schwarz1962,Kiefer1963,Lorden1976}.
In particular,  the ``indifference zone'' condition assumes that the null and alternative hypotheses are separated in the sense that the Kullback-Leibler divergence between the two hypotheses is positive, and if the true model parameter is in between the two hypotheses, then it is considered to be indifference for selecting the null and alternative hypothesis.
For example, for any $\delta>0,\kappa>0$, the set
\begin{equation}\label{eq:w-example}
     W = \{\ttt : \norm{\ttt}\leq \kappa \text{ and } \forall i\neq j \text{ such that }|\theta_i-\theta_j|\geq\delta \}
\end{equation}  satisfies Assumptions \ref{assump:compact}, \ref{assump:reg-w} and \ref{assump:separate}.  Assumption~\ref{assump:prior} requires the prior distribution to have a positive density function (bounded from zero) over the support. For instance, for the set $W$ described in \eqref{eq:w-example}, the uniform prior over $W$ satisfies the Assumption~\ref{assump:prior}.  
It is worthwhile to note that these technical assumptions are mainly for the theoretical development, while the proposed adaptive ranking policies are applicable in practice regardless of the conditions on $W$. 


Recall the definition of $D(\ttt)$ in \eqref{eq:obj}.
We further define
\begin{equation}
	t_c(\ttt)=\frac{|\log c|}{D(\ttt)}.
\end{equation}
Note that under the Assumption~\ref{assump:separate}, $t_c(\ttt)$ is always finite.
We first present a lower bound on the minimal Bayes risk ${V}^*_c(\rho)$ defined in \eqref{eq:V*}.
\begin{theorem}\label{thm:lb}
	Under Assumptions~\ref{assump:compact}-\ref{assump:prior}, we have
\begin{equation}
		\liminf_{c\to 0}\frac{{V}^*_c(\rho)}{c\EE t_c(\Theta)}\geq 1,
	\end{equation}
  where $\EE t_c(\Theta)= \int_W t_c(\ttt)\rho(\ttt)d\ttt$.
\end{theorem}\label{eq:lb}
Recall the definition in \eqref{eq:asymptotic-optimal} that a policy $\pi$ is said to be asymptotically optimal if $V_c(\pi,\rho)=(1+o(1))V_c^*(\rho)$ as $c\to 0$. Thus, to show a policy $\pi$ is indeed asymptotically optimal, we only need to show that $V_c(\pi,\rho)=(1+o(1))c\EE t_c(\Theta)$ as $c\to 0$, according to Theorem \ref{thm:lb}.
We proceed to show that the proposed sequential ranking method is asymptotically optimal. In Section~\ref{subsec:policy}, we propose two  policies $\pi_1=(A_p,T_1,R)$, $\pi_2=(A_p,T_2,R)$. Their risks consist of two parts,  the expected Kendall's tau and the expected sample size.
We start with some general upper bounds on the expected Kendall's tau for a class of pair selection schemes. {For the development of the upper bound, we further make the following two assumptions.}
\begin{assump}\label{assump:randomized}
There exists a positive constant $\delta_0$, such that
$$\min_{t,(i,j)}\lambda_t^{i,j}\geq |\log c|^{-\frac{1}{2}+\delta_0}$$
almost surely.
\end{assump}
{
\begin{assump}\label{assump:identifiability}
	For each $\ttt,\ttt'\in W$ and $\ttt\neq\ttt'$, there exists $a\in\mathcal{A}$ such that
	$
	f_{\ttt}^a(\cdot)\neq f_{\ttt'}^a(\cdot).
	$
\end{assump}}
\begin{theorem}\label{thm:error}
Under Assumption~\ref{assump:compact}-\ref{assump:prior} and Assumption~\ref{assump:identifiability}, we consider a policy $\pi_i=(A,T_i,{R})$ ($i=1,2$), where $A$ is a pair selection rule satisfying Assumption~\ref{assump:randomized} (not necessarily the proposed $\varepsilon$-greedy selection rule) and $R=\{R_{i,j}\}$.
Then,
 $$
	\EE L_K(\{R_{i,j}\})=O(c) \mbox{ for } i=1,2.
	$$
\end{theorem}
 We proceed to an upper bound on the expected sample size. The next assumption is needed on the selection scheme.
\begin{assump}\label{assump:optimal}
$$\lim_{c \to 0}  \sup_n \sum_{(i,j)}| \hat \lambda_n^{i,j}-\lambda_n^{i,j}| = 0,$$
where $\hat \llll_n = (\hat \lambda_n^{i,j})$ is defined in \eqref{eq:policy} and $\llll_n=(\lambda_n^{i,j})$ is the policy adopted at step $n$.
In other words, the policy $\hat \llll_n$ is adopted with probability $1-o(1)$ as $c\to 0$ at each step $n$.
\end{assump}
\begin{theorem}\label{thm:stopping-time}
	Under Assumption~\ref{assump:compact}-\ref{assump:prior} and Assumption~\ref{assump:identifiability}, we consider the policy $\pi=(A,T_i,{R})$ ($i=1,2$). If the pair assignment rule $A$ satisfies Assumption~\ref{assump:randomized} and~\ref{assump:optimal}, then
	$$
	\limsup_{c\to 0}\frac{\EE T_i}{\EE t_c(\Theta)} \leq 1  ~~(i=1,2).
	$$ 
\end{theorem}
Assumption~\ref{assump:identifiability} requires the identifiability of the model, which is critical for the consistency of the MLE.
Assumptions \ref{assump:randomized} and~\ref{assump:optimal} are assumptions on  the adaptive pair selection rule.
In particular, \ref{assump:randomized} requires that the selection rule explores every pair sufficiently,
 which is crucial for deriving the consistency of MLE. See below Lemma~\ref{lemma:mle} for the dependence of the deviation rate of MLE on the randomness of assignment rule. Assumption~\ref{assump:optimal} requires that $\hat \llll_n$ in \eqref{eq:policy}  is adopted with high probability, which is crucial for a sequential procedure to attain the asymptotic lower bound in Theorem~\ref{thm:lb}.

It is straightforward to see that if we choose the parameter $p$ in Algorithm \ref{algo:seq} such that  $p\geq |\log c|^{-\frac{1}{2}+\delta_0}$ and $p=o(1)$ as $c\to 0$, then the selection rule $A_p$ satisfies Assumption~\ref{assump:randomized} and \ref{assump:optimal}. Thus,
Theorems~\ref{thm:error} and \ref{thm:stopping-time} hold for $A=A_p$ when $p$ is appropriately chosen.
Combining this with the asymptotic lower bound on the minimal Bayes risk in Theorem~\ref{thm:lb}, and noticing that $\lim\limits_{c\to 0}\E t_c(\Theta)=\infty$, we  arrive at the asymptotic optimality of the proposed policies.
\begin{coro}\label{cor:opt}
	Under Assumption~\ref{assump:compact}-\ref{assump:prior}, and Assumption~\ref{assump:identifiability}, if we choose $p\propto |\log c|^{-\frac{1}{2}+\delta_0}$ for some $\delta_0$ satisfying $0<\delta_0<\frac{1}{2}$, then $\pi_i=(A_p,T_i,R)$, $i=1,2$ are asymptotically optimal policies.
\end{coro}
\paragraph{Consistency of MLE}
An auxiliary result obtained  in deriving the upper bound for the expected sample size is the following exponential bound for the MLE over a moving time window.
\begin{lemma}\label{lemma:mle}
	Let $m\geq n$ and $\varepsilon_{\lambda}= \min_{1\leq t\leq m,(i,j)}
	\lambda_t^{i,j}
	$ and $\varepsilon_1>0$. Then, for $n,m$ such that $n \varepsilon_{\lambda}^2\varepsilon_1^4\to\infty$, we have
	\begin{equation*}
		\PP_{\ttt}\Big(
		\sup_{n\leq t\leq m}\norm{\hat{\ttt}^{(t)}-\ttt}\geq \varepsilon_1
		\Big)\leq e^{-\Omega(n \epsilon_{\lambda}^2\varepsilon_1^4)}\times {O(m^{K}),}
	\end{equation*}
	where we denote $\PP_{\ttt}(\cdot)$  the conditional probability $\PP(\cdot|\Theta=\ttt)$.
\end{lemma}
The proof is provided in the supplementary material. From the above lemma, we can derive exponential upper bounds concerning the uniform consistency of $\hat{\ttt}^{(t)}$. In particular, if we let $\varepsilon_1$ be a fixed positive constant and $\varepsilon_{\lambda}^2\gg m^{-1}\log m $ as $m\to\infty$, then 
we can show
$\sup_{ t\geq n }\norm{\hat{\ttt}^{(t)}-\ttt}\to 0$ in probability as $n\to\infty$.

\subsection{Remarks}\label{subsec:discusion}

In this section, we provide some intuitions on the proposed policy as well as an efficient optimization algorithm for solving \eqref{eq:saddle_obj}.

\subsubsection{Intuitions}\label{sec:connections}
We provide some intuitions on the proposed stopping times \eqref{eq:t-1} and \eqref{eq:t-2} and MLE based decision rule on the inferred ranking.

{For the classic composite versus composite hypothesis testing problem with a zero-one loss without adaptive selection,
\cite{Schwarz1962}
 show that an asymptotic optimal stopping rule is the first passage time that the posterior error probability falls below a threshold $c$. Motivated by this, we consider a stopping rule decided by the posterior Kendall's tau.} To this end, let us first consider the minimization of posterior Kendall's tau in \eqref{eq:loss-k} under a fixed selection rule. Recall that $\Theta$ denotes the latent scores
with prior $\rho(\ttt)$.
One can define its posterior distribution after collecting $n$ comparison results $X_1,\ldots, X_n$.
Let $d^{(n)}=\{d_{i, j}^{(n)} \in \{0, 1\}, i < j \}$ be the pairwise decisions that minimize the expected value of $L_K$ with respect to the posterior distribution of $\Theta$:
\begin{equation}\label{eq:bayes_opt}
   d^{(n)} = \argmin_{R_{i,j} \in \{0,1\}, i<j }  \E\left( L_K(\{R_{i,j}\}) | \mathcal{F}_n \right).
\end{equation}
Note that in \eqref{eq:bayes_opt}, we do not require that the pairwise decisions $\{R_{i,j}\}_{i<j}$ form a global ranking.
Therefore, the above minimization problem can be solved separately for each $R_{i,j}$, for which
the optimal decision in \eqref{eq:bayes_opt} has the following form,
\begin{equation}\label{eq:pair_decision}
		d^{(n)}_{i,j}
   =\begin{cases}
	1 &\mbox{ if } \qquad \PP(\Theta_i>\Theta_j|\mathcal{F}_n) > \frac{1}{2};\\
	0 &\mbox{ otherwise}.
	\end{cases}	
\end{equation}
As we mentioned, a natural stopping time is to stop when the posterior Kendall's tau is below the cost $c$ of comparing one extra pair, i.e.,
 \begin{equation}\label{eq:t-3}
	T_3=\inf_{n > 1}\Big\{\E\Big[
	\sum_{i<j} I({\Theta_i>\Theta_j)}(1-d^{(n)}_{i,j}) + I(\Theta_i<\Theta_j)d^{(n)}_{i,j} |\mathcal{F}_n
	\Big]\leq c
	\Big\},
\end{equation}
By plugging $d_{i,j}^{(n)}$ in \eqref{eq:bayes_opt} into \eqref{eq:t-3}, we have
\begin{equation}\label{eq:T_1}
	T_3=\inf_{n>1}\Big\{
	\sum_{i<j} \min\{
	\PP(\Theta_i>\Theta_j|\mathcal{F}_n),1 - \PP(\Theta_i>\Theta_j|\mathcal{F}_n)
	\}\leq c\Big\}.
\end{equation}
However, the posterior probability $\PP(\Theta_i>\Theta_j|\mathcal{F}_n)$ is very complicated and thus the decision rule $d^{(n)}_{i,j}$ and $T_3$ cannot be directly computed. Therefore, we consider an approximation of the posterior probability.

Recall the definition  $W_{i,j}=\{ \ttt: \theta_i\geq\theta_j  \}\cap W$. Heuristically, { if the data are generated given parameter  $\ttt$ satisfying $\theta_i<\theta_j$}, the posterior probability $\PP(\Theta_i>\Theta_j|\mathcal{F}_n)$ has the following approximation when $n$ is large,
\begin{equation}\label{eq:approx}
\begin{aligned}
	\PP(\Theta_i>\Theta_j|\mathcal{F}_n) 
& \approx  \exp\left\{ \max\limits_{\ttt \in W_{i,j}} l_n(\ttt)-
\max_{\ttt \in W} l_n(\ttt) \right\},
\end{aligned}
\end{equation}
which is a standard approximation that has been used in the derivation of Bayesian information criterion \cite{schwarz1978}. Similarly, we approximate $1-\PP(\Theta_i>\Theta_j|\mathcal{F}_n)$ by  $\exp\big\{\max_{\ttt \in W_{j,i}} l_n(\ttt)-
\max_{\ttt \in W} l_n(\ttt) \big\}$.
By plugging the above approximations into \eqref{eq:T_1}, we obtain {a stopping rule
\begin{equation}\label{eq:approximation}
\inf\Big\{n > 1 : \sum_{(i,j) \in \mathcal{A}}
	    \exp\{-|\sup_{\tilde{\ttt}\in W_{i,j}}l_n(\tilde{\ttt})-\sup_{\ttt\in
        W_{j,i}}l_n(\ttt) | \}\leq c\Big\},	
\end{equation}
which is similar to $T_1$   defined in \eqref{eq:t-1}. The only difference is that \eqref{eq:t-1} adopts the  threshold  $e^{-h(c)}$ with $h(c)=|\log c| (1+ |\log c|^{-\alpha})$, while \eqref{eq:approximation} has a threshold
$c = e^{-|\log c|}$.
Note that $|\log c|^{-\alpha}$ is a $o(1)$ term when $c$ converges to $0$. 
The threshold $h(c)$ in \eqref{eq:t-1} is chosen slightly larger than $|\log c|$  for technical considerations (see Theorem~\ref{thm:error}).
}
If we further approximate the summation in \eqref{eq:approximation} by the maximization, a similar form of the stopping time $T_2$ is obtained. 
Roughly speaking, according to $T_1$ (or  $T_2$), the procedure stops when sufficient amount of information has been accumulated to distinguish all the pairs.

{Now we proceed to the decision rule. Note that when $n$ is large, the MLE $\hat{\ttt}^{(n)}$ is close to the true model parameter $\ttt$. We also note that $\PP(\Theta_i>\Theta_j|\mathcal{F}_n)\approx I(\theta_i>\theta_j)$ for large $n$. Combining this approximation with \eqref{eq:pair_decision}, we obtain an approximated decision rule $\tilde{d}^{(n)}_{i,j}\approx I(\hat{\theta}_{n,i}>\hat{\theta}_{n,j})$. It is straightforward to see that $\tilde{d}^{(n)}_{i,j}$ is the binary decision converted from the inferred ranking from MLE $r(\hat{\ttt}^{(n)})$, i.e., $\tilde{d}^{(n)}_{i,j}=1$ if and only if item $i$ is ranked higher than item $j$ according to $r(\hat{\ttt}^{(n)})$.
}

\subsubsection{Optimization in Algorithm~\ref{algo:seq}}
\label{sec:comp}

We adopt the  mirror descent algorithm (see, e.g., \cite{Beck:03}), as described in Algorithm~\ref{algo:mirror}, for solving
the optimization problem in \eqref{eq:saddle_obj}, i.e.,
\begin{equation}\label{eq:obj1}
 \argmax_{\llll\in\Delta} \min_{\tilde{\ttt}\in W: r(\tilde{\ttt})\neq r(\ttt)} \sum_{(i,j)}\lambda^{i,j}D^{i,j}(\ttt\|\tilde{\ttt}).
\end{equation}

\begin{algorithm}[!ht]
\begin{flushleft}
\caption{Mirror Descent Algorithm for Solving Eq. (\ref{eq:saddle_obj})}
\textbf{Input}: The MLE estimator $\ttt$ and total number of iterations $m$.

\textbf{Initialization}:  A starting point $\llll^0 \in \Delta$ and a constant $c_0 > 0$.


\textbf{Iterate} For $t=1,2,\ldots, m$:
 \begin{enumerate}
   \item Compute the maximizer:
   $$\tilde{\ttt}^0(\llll^{t-1}) \in \argmax_{\tilde{\ttt}\in W: r(\tilde{\ttt})\neq r(\ttt)} -\sum_{(i,j)}\lambda^{i,j,{t-1}}D^{i,j}(\ttt\|\tilde{\ttt})$$ 
   \item Compute the sub-gradient  $\mathbf{g}(\llll^{t-1})$ where $\mathbf{g}(\llll^{t-1})_{i,j}= -D^{i,j}(\ttt\|\tilde{\ttt}^0(\llll^{t-1}))$
   \item Update for $\llll^{t}$:
   \begin{equation}\label{eq:update}
   \llll^{t}=\argmin_{\llll \in \Delta} \left\{ \eta_t \langle \mathbf{g}(\llll^{t-1}), \llll \rangle + D(\llll \| \llll^{t-1})  \right\},
   \end{equation}
   where $\eta_t= \frac{c_0}{\sqrt{t}}$ and $D(\llll \| \llll^{t-1})$ is the KL divergence between $\llll$ and $\llll^{t-1}$, i.e.,  $D(\llll \| \llll^{t-1})=\sum_{i,j} \lambda_{i,j} \log \frac{\lambda^{i,j}}{\lambda^{i,j,{t-1}}}$
 \end{enumerate}
\textbf{Output}: The solution $\widehat{\llll}=\frac{1}{m}\sum_{t=1}^m \llll^{t}$.
 \label{algo:mirror}
\end{flushleft}
\end{algorithm}

We now elaborate steps of Algorithm~\ref{algo:mirror}. We first consider the inner optimization problem
\begin{equation}\label{eq:inner}
\tilde{\ttt}^0(\llll ) \in \argmax_{\tilde{\ttt}\in W: r(\tilde{\ttt})\neq r(\ttt)} -\sum_{(i,j)}\lambda^{i,j}D^{i,j}(\ttt\|\tilde{\ttt}),
\end{equation}
in step 1 of Algorithm~\ref{algo:mirror}.
For almost all the popular comparison models,   the objective function $-\sum_{(i,j)}\lambda^{i,j}D^{i,j}(\ttt\|\tilde{\ttt})$ is smooth in $\tilde \ttt$. 
Moreover, the objective function is also concave in $\tilde{\ttt}$ for comparison models in an exponential family form (e.g., the BTL model in \eqref{eq:BTL}).
When the support $\{\tilde{\ttt}\in W: r(\tilde{\ttt})\neq r(\ttt)\}$ can be written as the union of a finite number of convex sets,  \eqref{eq:inner} can be obtained by solving finite optimization problems, each with a smooth objective function constrained in a convex set.
For moderately large $K$,
such problems can typically be solved well by standard numerical solvers.
Therefore, from now on,  we assume that the inner optimization problem can be solved.

We then discuss the outer optimization problem
\begin{equation}\label{eq:h}
  \min_{\llll\in\triangle} h(\llll), ~ h(\llll)= \max_{\tilde{\ttt}\in W: r(\tilde{\ttt})\neq r(\ttt)} \phi(\llll, \tilde{\ttt}), ~ \phi(\llll,\tilde{\ttt})=-\sum_{(i,j)}\lambda^{i,j}D^{i,j}(\ttt\|\tilde{\ttt}).
\end{equation}
When $\phi(\llll, \tilde{\ttt})$ is a continuous and bounded function and the set $W$ is compact,
further noting that $\phi(\llll,\tilde{\ttt})$ is convex in $\llll$ for every $\tilde{\ttt}$,
 $h(\llll)$ is a convex function in $\llll$, by the Danskin's Theorem (see Proposition B.25 in \cite{Ber:99}).
Moreover, for a given $\llll$, let $\tilde{\ttt}^0(\llll) \in \argmax_{\tilde{\ttt}\in W: r(\tilde{\ttt})\neq r(\ttt)} \phi(\llll, \tilde{\ttt})$ be one of the maximizers. Then, by Danskin's theorem,  $\mathbf{g}(\llll)$ with $\mathbf{g}(\llll)_{i,j}= -D^{i,j}(\ttt\|\tilde{\ttt}^0(\llll))$ is a sub-gradient of $h(\llll)$, as used in step 2 of Algorithm~\ref{algo:mirror}.

Finally, \eqref{eq:update} in step 3 of the algorithm has a closed-form solution, obtained by by writing down the KKT condition. That is,
\[
  \lambda^{{i,j},t}=\frac{1}{C}\lambda^{i,j,{t-1}} \exp\left(-\eta_t \mathbf{g}(\llll^{t-1})_{i,j} \right),
\]
\sloppy where $\lambda^{{i,j},t}$ is the $(i,j)$-th component of $\llll^{t}$ and the normalization constant $C=\sum_{i,j} \lambda^{i,j,t-1} \exp\left(-\eta_t \mathbf{g}(\llll^{t-1})_{i,j} \right)$.
%

Under the mild conditions as above and assuming that the inner optimization can be solved,
this mirror descent algorithm is guaranteed to converge to the solution of the optimization program
at the rate of $O\left(\sqrt{{1}/{t}}\right)$, i.e.,  $h(\hat{\llll})- \min_{\llll \in \Delta }h(\llll)=O\left(\sqrt{{1}/{t}}\right)$ (see, e.g., \cite{Beck:03} or Theorem 4.2 from \cite{Bubeck:15:Opt}).

In practice,  support $W$ of the prior distribution $\rho(\cdot)$ maybe unknown. In this case, we may choose
\begin{equation}\label{eq:missupp}
W = \cup_{(i,j)} W_{i, j} \mbox{~and~} W_{i,j}=\{\ttt: \theta_i \geq \theta_j\}\cap \{\ttt: |\theta_i|\leq M, 2\leq i\leq K\}
\end{equation}
in the design of sequential ranking policy for some positive constant $M$.
With this mis-specified support of $\rho(\cdot)$, the resulting policy may not achieve the asymptotic lower bound of the Bayes risk presented in Theorem~\ref{thm:lb}, due to the incomplete information. On the other hand, the Bayes risk of the resulting ranking procedure can still achieve the same order of the minimal Bayes risk as $c\to 0$. That is, $\limsup_{c\rightarrow 0}{V_c(\rho,\pi)}/{V^*_c(\rho)}$ may be finite but greater than $1$.%

\section{Simulation Study}
\label{sec:simu}
\subsection{Study I: Asymptotic Optimality}

\begin{figure}
  \centering
  \includegraphics[width=.6\linewidth]{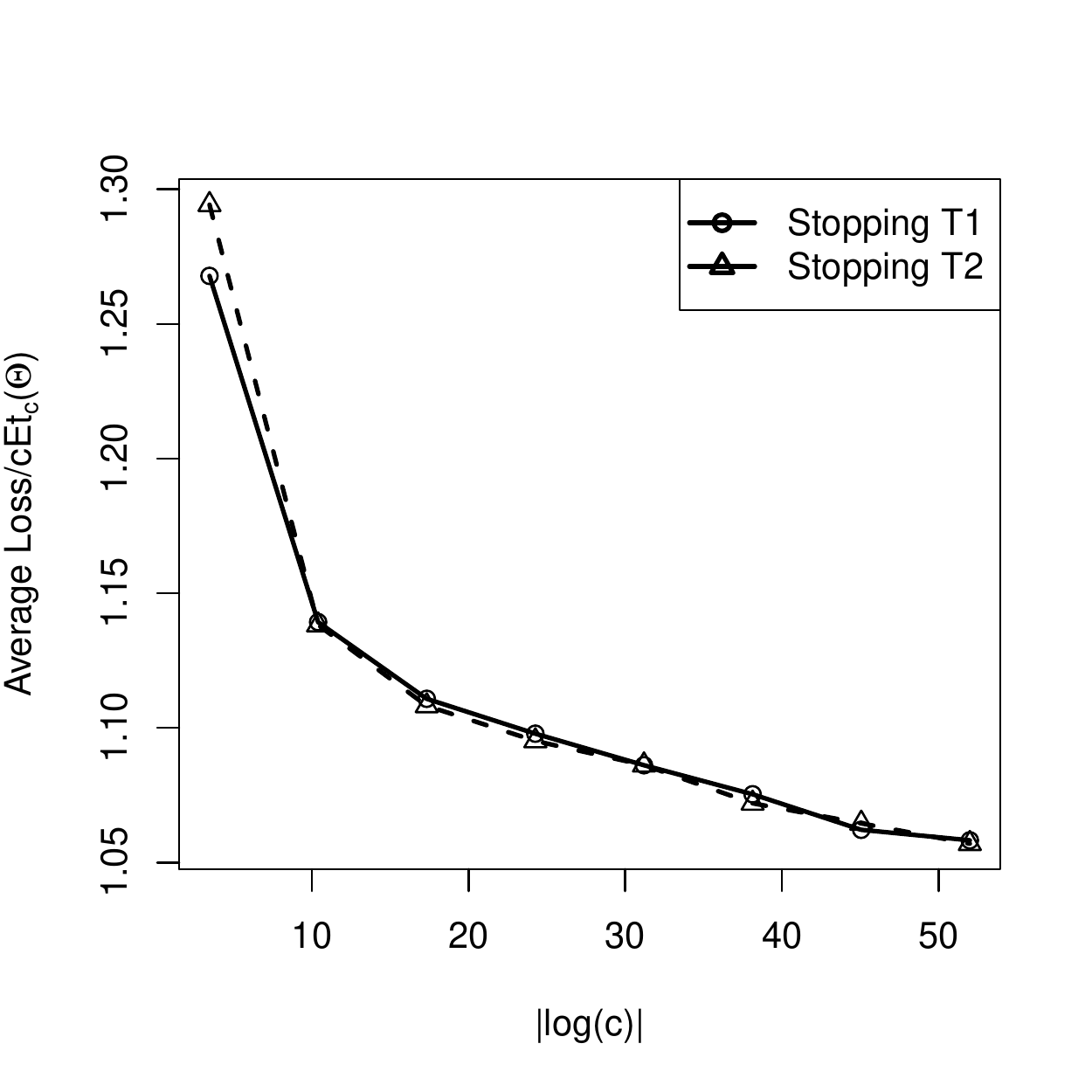}
\caption{Results of  Simulation Study I.}
\label{fig:studyI}
\vspace{-5mm}
\end{figure}

We first provide a simulation study to check the main theoretical result in Section~\ref{subsec:optimal}.
We consider $K = 3$ items and
$$W = \{\ttt = (\theta_2, \theta_3): \Vert \ttt\Vert_{\infty} \leq  2, \text{ and } |\theta_i-\theta_j|\geq 0.4, i\neq j, i,j = 1, 2, 3\},$$
where $\theta_1 = 0$ according to our assumption and $\Vert \ttt \Vert_{\infty} = \max\{\vert \theta_i\vert: i = 2, 3\}$ is the supremum norm.
Latent score $\Theta$ follows a uniform distribution on $W$. In addition,
a range of values of $c$ are considered, including $2^{-5}$, $2^{-15}$, $2^{-25}$, ..., $2^{-75}$.
In this study, the support $W$ is assumed to be known when applying Algorithm~\ref{algo:seq}.
Results based on the two proposed stopping rules $T_1$ and $T_2$ are shown in  Figure~\ref{fig:studyI}, where
the $x$-axis represents $\vert \log c\vert$ and the $y$-axis represents the ratio between the average loss $\bar V$ and $c\E t_c(\Theta)$.
According to Figure~\ref{fig:studyI}, for each stopping rule, the ratios are above one and decreases as
$c$ decreases
(i.e., $\vert\log c\vert$ increases). They tend to decay to 1 as $c$ converges to 0.

\subsection{Study II: Comparison}\label{subsec:compare}

We then compare the proposed methods with (1) an algorithm that has randomized selection and fixed-length stopping
and (2) an algorithm that selects comparison pair based on Wald statistic with  fixed-length stopping. More precisely, at each step $n$, the Wald-statistic based algorithm computes the MLE $\hat \ttt^{(n-1)}$ and its asymptotic variance based on the observed Fisher information. Then for each pair $i$ and $j$, we compute the standard error of
$\hat \theta_i^{(n-1)} - \hat \theta_j^{(n-1)}$ by delta method, denoted by $\hat \sigma_{ij}^{(n-1)}$.
The Wald statistic for testing $\theta_i = \theta_j$ versus  $\theta_i \neq \theta_j$
 is defined as
 $$Z_{ij}^{(n-1)} \triangleq (\hat \theta_i^{(n-1)} - \hat \theta_j^{(n-1)}) /\hat \sigma_{ij}^{(n-1)}.$$
Roughly speaking, the larger the absolute value of the Wald statistic, the easier to distinguish the two items. Therefore, the
algorithm chooses the pair with the smallest $\vert Z_{ij}^{(n-1)}\vert$ for comparison in the next stage.

\begin{figure}[!t]
\centering
\begin{subfigure}{.5\textwidth}
  \centering
  \includegraphics[width=0.9\linewidth]{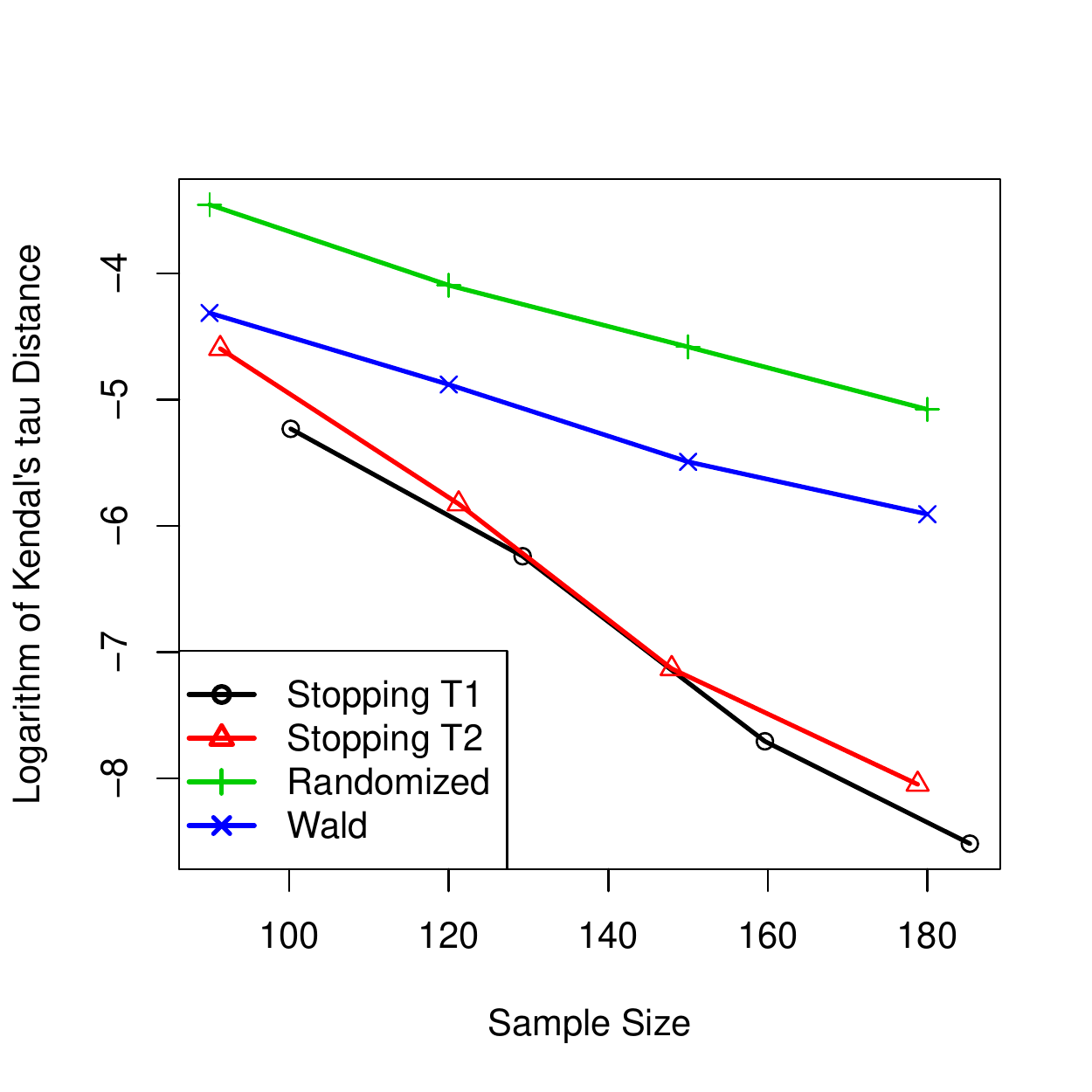}
  \caption{$K = 3$.}
  \label{fig:sub1}
\end{subfigure}%
\begin{subfigure}{.5\textwidth}
  \centering
  \includegraphics[width=.9\linewidth]{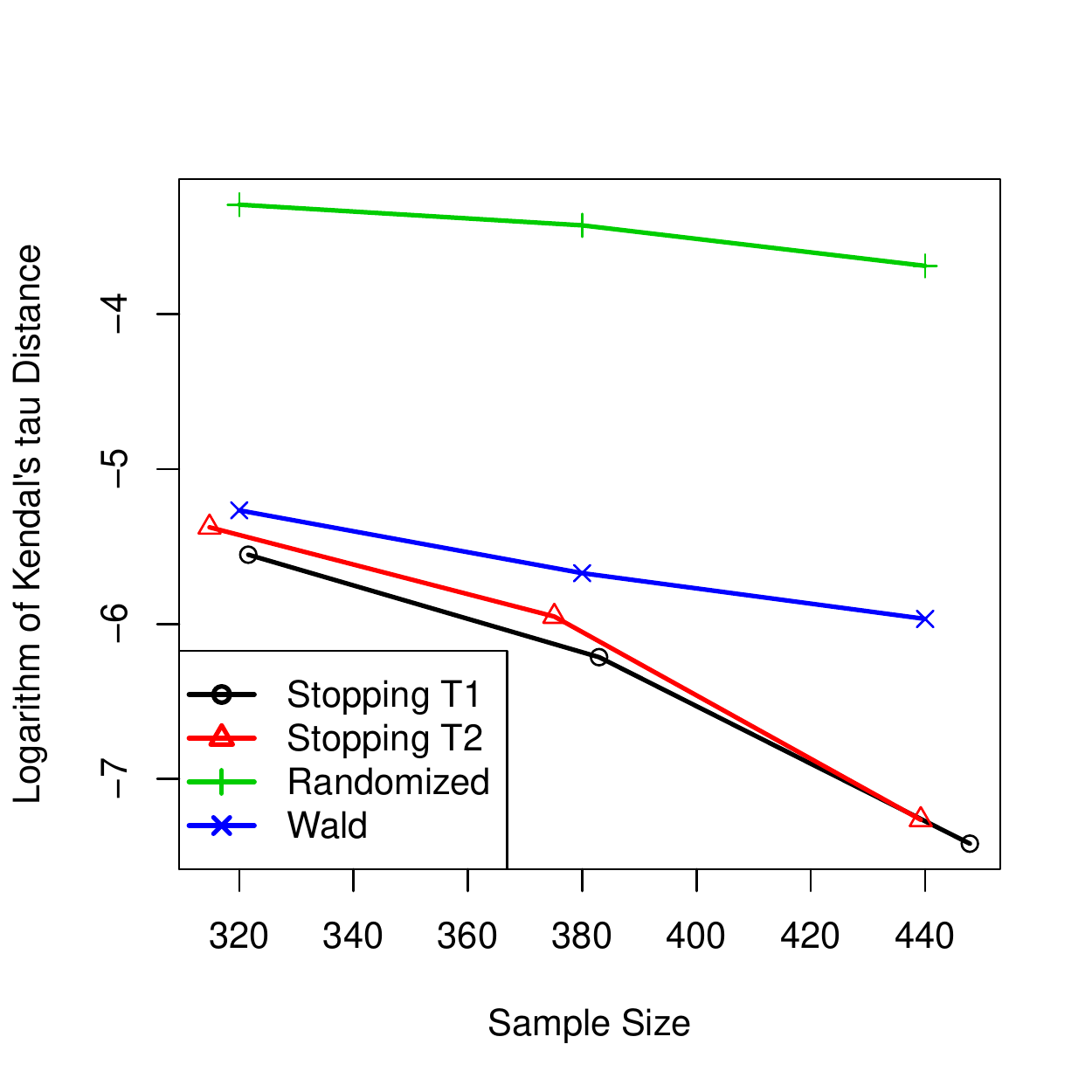}
  \caption{$K = 4$.}
  \label{fig:sub2}
\end{subfigure}
\caption{Results of Simulation Study II.}
\label{fig:studyII}
\vspace{-5mm}
\end{figure}

We consider two settings, with $K = 3$ and $K = 4$.  When $K = 3$, the same setting as in Study I is used. When $K = 4$, we let
$$W = \{\ttt = (\theta_2, ..., \theta_4): \Vert \ttt \Vert_{\infty} \leq 4, \text{ and } |\theta_i-\theta_j|\geq 0.2, i\neq j, i,j = 1, ..., 4\}.$$
Latent score $\Theta$ follows a uniform distribution over $W$. In this study, the support $W$ is assumed to be unknown when applying Algorithm~\ref{algo:seq}. In particular, we choose
$W$ as specified in \eqref{eq:missupp}, with $M = 5$.
Results are shown in Figure~\ref{fig:studyII}. For the proposed two methods, each point corresponds to a value of $c$ and for the two competitors,
each point corresponds to a given sample size. The $x$-axis represents the average of sample size and $y$-axis represents the average of Kendall's tau distance.
According to the results,
the proposed two methods perform similarly and both substantially outperform the randomized and Wald statistic based algorithms. In addition, the Wald statistic based algorithm performs significantly better than the randomized one.

\section{Concluding Remarks}
\label{sec:conclusion}

In this paper, we consider the sequential design of rank aggregation with adaptive pairwise comparison.
This problem is not only of practical importance due to its wide applications in fields such as
psychology,  politics, marketing, and sports, but also of theoretical significance in sequential analysis.
Due to the more complex structure of the ranking problem than hypothesis testing problems, no existing sequential
analysis framework is suitable.
We formulate the problem under
a Bayesian decision framework
and develop asymptotically optimal policies. Comparing to the existing Bayesian sequential hypothesis testing problems,
the problem solved in this paper is technically more challenging due to the more structured risk function.
Novel technical tools are developed to solve this problem, which are of separate theoretical interest in solving complex sequential design problems.

The current work may be extended in several directions. First, an even larger class of comparison models may be considered.
The models considered in the current paper
all assume the judges being homogeneous, i.e., the
comparison outcome does not depend on who the judge is.
It is of interest to consider the heterogeneity of the judges by
incorporating judge-specific random effects
into the comparison models and develop corresponding sequential designs.
Second, different risk structures will be incorporated into the
sequential ranking designs to account for practical needs in different applications.
For example, we will consider other metrics for assessing the
ranking accuracy (e.g. based on the accuracy of identifying the set of top $K$ items)
and non-uniform costs for different judges.

%

\section{Proof of Theorems}\label{sec:proof}
In this section, we present the proofs of Theorem~\ref{thm:lb}-\ref{thm:stopping-time}. The proof for lemmas are delayed in the supplementary material. 
\sloppy Throughout the proof, we will use the constants $\delta_{\rho}=\inf_{\ttt\in W}\rho(\ttt)>0$ and $\sup_{\ttt\in W, x, a\in \mathcal{A}}|\nabla f^{a}_{\ttt}(x)|\leq \kappa_{0}$. According to Assumptions~\ref{assump:prior} and \ref{assump:f}, these two constants are finite.

\subsection{Proof for Theorem~\ref{thm:lb}}\label{sec:proof-lb}
Let $\varepsilon= c|\log c|^2$.
For an arbitrary policy $\pi=(A,T,R)$ and a prior probability density function $\rho$, there are two possibilities: either $\E L_K(R)\geq \varepsilon $ or $\E L_K(R) <\varepsilon$. For the first case, we can see ${V}(\rho,\pi)\geq \varepsilon\geq (1+o(1))c\E t_c(\Theta)$.
For the second case, we have 
$$
V(\pi,\rho)= \E L_K(R)+ c\E T \geq c\E T.
$$
Therefore, to prove the theorem it is sufficient to show that
$$
\liminf_{c\to 0} \frac{c\E T}{c\E t_c(\Theta)}\geq 1
$$
or, equivalently, for each $\delta>0$ there exists a positive constant $c_0>0$ such that for $c<c_0$,
$$
\E T \geq (1-\delta) \E t_c(\Theta).
$$
Let $t_{c,\delta}(\ttt)=(1-2\delta/3)t_c(\ttt)$  for each $\delta>0$. 
Then we arrive at a lower bound
\begin{align*}
	\EE T &\geq \EE[T; T>t_{c,\delta}(\Theta)]\\
	 &\geq \int \rho(\ttt) t_{c,\delta}(\ttt) \PP_{\ttt}(T>t_{c,\delta}(\ttt))d\ttt\\
	 &= \EE t_{c,\delta}(\Theta) - \int \rho(\ttt) t_{c,\delta}(\ttt) \PP_{\ttt}(T\leq t_{c,\delta}(\ttt))d\ttt\\
	 &\geq \EE t_{c,\delta}(\Theta) - t_{\max} \PP(T\leq t_{c,\delta}(\Theta)),
\end{align*}
where we define $t_{\max}=\max_{\ttt\in W}t_{c,\delta}(\ttt)$
 and recall that  $\PP_{\ttt}$ represents for the conditional probability $\PP(\cdot|\Theta=\ttt)$.
 According to Assumption~\ref{assump:separate} we have $t_{\max}= O(|\log c|)=O(\E t_c(\Theta))$.
Therefore, it is sufficient to show
\begin{equation}
	\PP(T\leq t_{c,\delta}(\Theta)) =o(1).
\end{equation}
We proceed to an upper bound for $\PP(T\leq t_{c,\delta}(\Theta))$.
We abuse the notation a little and write
$
U_r=\{\ttt:r(\ttt)=r \}
$,  the set of parameters that gives the rank $r$. Then, we have
\begin{equation}\label{eq:lowerbound-sum}
\begin{split}
	\PP(T\leq t_{c,\delta}(\Theta))=&\sum_{r\in P_K} \PP(T\leq t_{c,\delta}(\Theta),\Theta\in U_r)\\
	\leq & O(1)\times \max_{r\in P_K}\PP(T\leq t_{c,\delta}(\Theta),\Theta\in U_r).
\end{split}
\end{equation}
We proceed to an upper bound for $\PP(T\leq t_{c,\delta}(\Theta),\Theta\in U_r)$ for each $r\in P_K$.
Define an event
\begin{equation}\label{eq:B}
	B=\Big\{
\frac{
\PP(\Theta\in U_r|\FF_T)
}{\max_{(i,j):W_{i,j}\cap U_r=\emptyset}\PP(\Theta\in W_{i,j}|\FF_T)}>\frac{c^{\frac{\delta
}{10}}}{\varepsilon}
\Big\},
\end{equation}
where $\FF_n=\sigma(X_1,...,X_n,a_1,...,a_n)$ denotes the $\sigma$-algebra generated by $X_1,...,X_n$ and $a_1,..,a_n$.
We split the probability
\begin{equation}
\begin{split}
	&\PP(T\leq t_{c,\delta}(\Theta),\Theta\in U_r)\\
	=&\PP(T\leq t_{c,\delta}(\Theta),\Theta\in U_r, B)+\PP(T\leq t_{c,\delta}(\Theta),\Theta\in U_r, B^c),
\end{split}
	\end{equation}
which can be bounded from above by
\begin{equation}\label{eq:two-terms}
	\PP(T\leq t_{c,\delta}(\Theta),\Theta\in U_r)\leq \PP(T\leq t_{c,\delta}(\Theta),\Theta\in U_r, B)
	+ \PP(\Theta\in U_r, B^c).
\end{equation}
We establish upper bounds for the two terms on the right-hand side of the above inequality separately.
The next lemma, whose proof is presented in the supplementary material, provides an upper bound for the second term.
\begin{lemma}\label{lemma:second-term}
For all $r\in P_K$, if $\E L_K(R)\leq \varepsilon$ then
	$$
	\PP(\Theta\in U_r, B^c)\leq (1+\frac{c^{\frac{\delta}{10}}}{\varepsilon})\varepsilon.
	$$
\end{lemma}
We proceed to the first term $\PP(T\leq t_{c,\delta}(\Theta),\Theta\in U_r, B)$ on the right-hand side of \eqref{eq:two-terms}.  Then,
\begin{equation}\label{eq:first-term-exp}
	\PP(T\leq t_{c,\delta}(\Theta),\Theta\in U_r, B)= \int_{U_r} \PP_{\ttt}(T\leq t_{c,\delta}(\ttt), B)\rho(\ttt)d\ttt.
\end{equation}
Recall the definition of the event $B$ in \eqref{eq:B}, we have
\begin{equation}
	B\cap\{T\leq t_{c,\delta}(\ttt)\}\subset \Big\{
\max_{1\leq t\leq t_{c,\delta}(\ttt)}\frac{\PP(\Theta\in U_r|\FF_{t})
}{\max_{W_{i,j}\cap U_r=\emptyset}\PP(\Theta\in W_{i,j}|\FF_{t})}>\frac{c^{\frac{\delta}{10}}}{\varepsilon}
	\Big\}.
\end{equation}
Consequently,
\begin{equation}\label{eq:first-term}
	\PP_{\ttt}\left(T\leq t_{c,\delta}(\ttt), B\right)\leq
	\PP_{\ttt}\Big(
\max_{1\leq t\leq t_{c,\delta}(\ttt)}\frac{\PP(\Theta\in U_r|\FF_{t})
}{\max_{W_{i,j}\cap U_r=\emptyset}\PP(\Theta\in W_{i,j}|\FF_{t})}>\frac{c^{\frac{\delta}{10}}}{\varepsilon}
	\Big).
\end{equation}
We proceed to an upper bound for the above display. For each $\ttt$, we define a random sequence $\{\ttt_t^*:1\leq t\leq t_{c,\delta}(\ttt) \}$ as follows.
\begin{equation}\label{eq:thetastar}
	\ttt^*_t= \arg\min_{\tilde{\ttt}\in W:r(\tilde{\ttt})\neq r(\ttt) }\sum_{n=1}^t\sum_{i,j} \lambda_n^{i,j} D^{i,j}(\ttt\|\tilde{\ttt}).
\end{equation}
Intuitively, $\ttt^*_t$ is the score parameter that is most difficult to distinguish from $\ttt$ at time $t$ among those that have different rank with $\ttt$, {given that item selection rules $\lambda_1, ..., \lambda_n$ have been adopted.}
We further choose the index process $(i^*_t, j^*_t)$ be such that $\ttt^*_t\in W_{i^*_t, j^*_t}$ but $\ttt\notin  W_{i^*_t, j^*_t}$. If there are multiple $(i,j)$'s satisfying this, then we choose $(i^*_t,j^*_t)$ arbitrarily from them.
From the definition, we know
$\ttt_t^*$ and  $(i^*_t, j^*_t)$ are adapt to $\sigma(\lambda_1,...,\lambda_t)$, and thus are adapt to $\mathcal F_{t-1}$.
We use the next lemma to transform the probability in \eqref{eq:first-term} to a probability based on a martingale parameterized by $\ttt$. 
\begin{lemma}\label{lemma:transform-first-term}
For each $\ttt'\in U_r$, define a martingale with respect to the filtration
$\{\FF_{n}:n\geq 1\}$ and probability measure $\PP_{\ttt}$
as follows, 
\begin{equation}\label{eq:mart}
 	M_t(\ttt')=l_t^{\vec{a}}(\ttt')-l_t^{\vec{a}}(\ttt^*_t)-\sum_{n=1}^t\sum_{(i,j)}\lambda_n^{i,j}D^{i,j}(\ttt\|\ttt^*_t) + \sum_{n=1}^t\sum_{(i,j)}\lambda_n^{i,j}D^{i,j}(\ttt\|\ttt'),
\end{equation}
where $l^{\vec{a}}_{t}(\ttt)= \log \prod_{i=1}^t f^{a_i}_{\ttt}(X_i)$.
Then there exists a positive constant $c_0>0$ such that for $0<c<c_0$, 
\begin{align}
	&\PP_{\ttt}\Big(
\max_{1\leq t\leq t_{c, \delta}(\ttt)}\frac{\PP(\Theta\in U_r|\FF_{t})
}{\max_{W_{i,j}\cap U_r=\emptyset}\PP(\Theta\in W_{i,j}|\FF_{t})}>\frac{c^{\frac{\delta}{10}}}{\varepsilon}\Big)\\
\leq&\PP_{\ttt}\Big(
\max_{1\leq t\leq t_{c, \delta}(\ttt),\ttt'\in U_r}M_t(\ttt')\geq \frac{\delta}{2}|\log c|
\Big).\label{eq:level-crossing-prob1}
\end{align}
\end{lemma}
According to the above lemma, to find an upper bound for \eqref{eq:first-term}, it is sufficient to find an upper bound for the right-hand side of \eqref{eq:level-crossing-prob1}, which
 is the probability that a stochastic process indexed by $\ttt'$ and $t$ goes above a certain level. In this paper, we will use the following two lemmas repeatedly to handle this type of level crossing probabilities. The first one is the Azuma-Hoeffding inequality proved by \cite{azuma1967weighted} and \cite{hoeffding1963probability}.
\begin{lemma}[Azuma-Hoeffding inequality]\label{lemma:hoeffding}
	Let $M_n$ be a martingale with respect to the filtration $\{\mathcal{F}_n:n=1,2,.. \}$. Let $X_n= M_n-M_{n-1}$. Assume that $X_n$ is bounded and $X_n\in[a_n,b_n]$ where $a_n$ and $b_n$ are deterministic constants. Then,  for each $t>0$ we have
	\begin{equation}
		\PP(\max_{1\leq m\leq n}M_m\geq t)\leq \exp\Big(-\frac{2t^2}{\sum_{i=1}^n(b_i-a_i)^2 }\Big).
	\end{equation}
\end{lemma}
The next lemma is the key lemma that allows us
to derive level crossing probability by aggregating marginal tail bounds of a random field. Its proof is given in the supplementary material. 
\begin{lemma}\label{lemma:level-crossing-prob}
	Let $\{\zeta(\ttt):\ttt\in U\}$ be a random field over a compact set $U\subset \RR^K$ that satisfies Assumption~\ref{assump:reg-w}.
  Let $\beta(\ttt,b)$ be defined as follows
	$$
	\beta(\ttt,b)=\PP( \zeta(\ttt)\geq b),
	$$
	where $\PP$ is a probability measure and we assume that $\zeta(\cdot)$ has continuous sample path almost surely under $\PP$.
	Assume that $\zeta(\cdot)$ has a Lipschitz continuous sample path in the sense that there exists a constant $\kappa_{L}$ such that for all $\theta,\theta'\in W$
	$$
	|\zeta(\ttt)-\zeta(\ttt')|\leq\kappa_{L} \|
	\ttt-\ttt'
	\| \text{ almost surely under } \PP.
	$$
	Then, we have that for all positive $\gamma$
	$$
	\PP\Big(
\max_{\theta\in W} \zeta(\ttt)\geq b \Big)\leq \int_{W} \beta(\ttt,b-\gamma) d\ttt \times \frac{\kappa_L^{K-1}}{\gamma^{K-1} \delta_b},
$$
where $\delta_b$ is the constant defined in Assumption~\ref{assump:reg-w}.
\end{lemma}
Set $n:=t_{c,\delta}(\ttt)$, $t:=\frac{\delta}{2}|\log c|-1$,  $M_n:=M_n(\ttt')$, and $a_n=-b_n:=2\max_{x,a\in\mathcal{A},\theta\in W}|\log f^a_{\theta,x}(x)|$
in Lemma~\ref{lemma:hoeffding}, we have for each $\ttt'$
\begin{equation}
			\PP_{\ttt}\left(\max_{1\leq n\leq t_{c,\delta}(\ttt)}M_n(\ttt')\geq \frac{\delta}{2}|\log c|-1\right)\leq \exp\Big(-\frac{2(\frac{\delta}{2}|\log c|-1)^2}{t_{c,\delta}(\ttt) a_1^2}\Big).
\end{equation}
According to Assumption~\ref{assump:compact} and \ref{assump:f},
we have $a_1<\infty$, and consequently,
\begin{equation}\label{eq:marginal-bound-1}
			\PP_{\ttt}\left(\max_{1\leq n\leq t_{c,\delta}(\ttt)}M_n(\ttt')\geq \frac{\delta}{2}|\log c|-1\right)\leq \exp\Big(-\Omega( \delta^2 |\log c|)
			\Big).	
\end{equation}
Note that for $\ttt',\tilde{\ttt}\in U_r$,
\begin{equation}
\begin{split}
	&	\max_{1\leq n\leq t_{c,\delta}(\ttt)} M_n(\ttt') - \max_{1\leq n\leq t_{c,\delta}(\ttt)} M_n(\tilde{\ttt})\\
\leq & \max_{1\leq n\leq t_{c,\delta}(\ttt)} |M_n(\ttt') - M_n(\tilde{\ttt})|\\
\leq &t_{c,\delta}(\ttt)\kappa_0 \Vert\ttt'-\tilde{\ttt}\Vert,
\end{split}
\end{equation}
where $\kappa_0=4\sup_{a\in\mathcal{A},\ttt'\in W,x}|\nabla \log f^{a}_{\ttt}(x) |<\infty$ denotes the Lipschitz constant of $M_1(\ttt')$.
Therefore, $M_n(\ttt')$ is a Lipschitz continuous random field in $\ttt'$.
The above display and \eqref{eq:marginal-bound-1}, together with Lemma~\ref{lemma:level-crossing-prob}, give
\begin{equation}
\begin{split}
	&\PP_{\ttt}\left(\max_{1\leq n\leq t_{c,\delta}(\ttt),\ttt'\in U_r}M_n(\ttt')\geq \frac{\delta}{2}|\log c|\right)\\
	\leq& \exp\Big(-\Omega( \delta^2 |\log c|)
			\Big) \text{meas}(U_r)\frac{t_{c,\delta}(\ttt)^{K-1}\kappa_0^{K-1}}{\delta_b}\\
			= & \exp\Big(-\Omega( \delta^2 |\log c|)
			\Big)\times O(|\log c|^{K-1}).
\end{split}			
\end{equation}
The above inequality and \eqref{eq:first-term-exp}, \eqref{eq:first-term},\eqref{eq:level-crossing-prob1} give
\begin{equation}
	\PP(T\leq t_{c,\delta}(\Theta),\Theta\in U_r, B)\leq
			 \exp\Big(-\Omega(\delta^2 |\log c|)
			\Big)\times O(|\log c|^{K-1}).
\end{equation}
Combine this with Lemma~\ref{lemma:second-term} and \eqref{eq:two-terms} we have
\begin{equation}
	\PP(T\leq t_{c,\delta}(\Theta),\Theta\in U_r)
 	\leq (1+\frac{c^{\frac{\delta}{10}}}{\varepsilon})\varepsilon + \exp\Big(-\Omega( \delta^2 |\log c|)
			\Big)\times O(|\log c|^{K-1}).
\end{equation}
Combine the above display with \eqref{eq:lowerbound-sum}, we have
\begin{equation}
	\PP(T\leq t_{c,\delta}(\Theta))
 	\leq O(1)\times\Big\{(1+\frac{c^{\frac{\delta}{10}}}{\varepsilon})\varepsilon + \exp\Big(-\Omega( \delta^2 |\log c|)
			\Big)\times O(|\log c|^{K-1})\Big\}.	
\end{equation}
Therefore, $\PP(T\leq t_{c,\delta}(\Theta))=o(1)$ as $c\to 0$. This completes the proof.
\subsection{Proof of Theorem~\ref{thm:error}}
We start with the stopping time $T_2$.
 With the decision rule ${D}$ defined in \eqref{eq:decision}, the expected Kendall's tau at the stopping time $T_2$ is
\begin{equation}\label{eq:kendal-upper}
\begin{split}
	\EE L_K({R}) = &
\EE \sum_{(i,j)} I(\Theta_i<\Theta_j)R_{i,j}\\
 =&\int_W \sum_{(i,j):\ttt\notin W_{j,i}} \PP_{\ttt}(\sup_{\tilde{\ttt}\in W_{j,i}}l_{T_2}(\tilde{\ttt})>\sup_{\ttt'\in W_{i,j}} l_{T_2}(\ttt'))\rho(\ttt)d\ttt\\
	=& \int_W \sum_{\ttt\notin W_{j,i}} \PP_{\ttt}\Big(\sup_{\tilde{\ttt}\in W_{j,i}}l_{T_2}(\tilde{\ttt})-\sup_{\ttt'\in W_{i,j}}l_{T_2}({\ttt}')> h(c)\Big)\rho(\ttt)d\ttt,
\end{split}
\end{equation}
where we write $l_t(\ttt)=\sum_{n=1}^t \log f^{a_n}_{\ttt}(X_n)$ as the log-likelihood function.
\eqref{eq:kendal-upper} is bounded from above by
\begin{equation}\label{eq:two-terms-split}
\begin{split}
		\EE L_K({R}) \leq &
	\sup_{\ttt\in W} \rho(\ttt) \times \text{meas}(W)\times \frac{K(K-1)}{2} \\
	&\times \sup_{\ttt\in W} \max_{(i,j):\ttt\notin W_{j,i}} \PP_{\ttt}\Big(\sup_{\tilde{\ttt}\in W_{j,i}}l_{T_2}(\tilde{\ttt})-l_{T_2}(\ttt)> h(c)\Big).
\end{split}
\end{equation}
To obtain the above inequality, we used the fact that $\sup_{\ttt'\in W_{i,j}}l_{T_2}({\ttt}')\geq l_{T_2}(\ttt)$ for $(i,j)$ such that $\ttt\notin W_{j,i}$.
We split the probability
	\begin{equation}\label{eq:two-terms-1}
	\begin{split}
			& \PP_{\ttt}\left(
		\sup_{\tilde{\ttt}\in W_{ji} }l_{T_2}(\tilde{\ttt}) - l_{T_2}(\ttt)> h(c)
		\right)\\
		\leq & \PP_{\ttt}\left(
		\sup_{\tilde{\ttt}\in W_{ji} } l_{T_2}(\tilde{\ttt}) - l_{T_2}(\ttt)> h(c); T_2\leq \tau
		\right) + \PP_{\ttt}(T_2\geq \tau).
	\end{split}
		\end{equation}
The second term on the right-hand side of the above display is controlled by the next lemma.
\begin{lemma}\label{lemma:t-bound}
	If $\tau=\Omega(|\log c|^{\tup})$
	 then for any selection rule satisfying the Assumption~\ref{assump:randomized}, we have
\begin{equation}
	\PP_{\ttt}(T_i\geq\tau)\leq c^2 ~~~~(i=1,2).
\end{equation}
\end{lemma}
We proceed to an upper bound of the first term on the right-hand side of \eqref{eq:two-terms-1}.
		Define a stopping time $T_2\wedge \tau=\min(T_2,\tau)$, then we have
		\begin{equation}
		\begin{split}
				&\PP_{\ttt}\left(
		\sup_{\tilde{\ttt}\in W_{ji} } l_{T_2}(\tilde{\ttt}) - l_{T_2}(\ttt)> h(c); T_2\leq \tau
		\right)\\
		\leq & \PP_{\ttt}\left(
		\sup_{\tilde{\ttt}\in W_{ji} } l_{T_2\wedge \tau}(\tilde{\ttt}) - l_{T_2\wedge \tau}(\ttt)> h(c)
		\right).
		\end{split}
			\end{equation}
Now we consider the random field $\eta(\tilde{\ttt})=l_{T_2\wedge \tau}(\tilde{\ttt}) - l_{T_2\wedge \tau}(\ttt)$ for $\tilde{\ttt}\in W_{ji}$. We proceed to an upper bound for
$
\PP_{\ttt}\left(
		\sup_{\tilde{\ttt}\in W_{ji} } \eta({\tilde{\ttt}})> h(c)
		\right)
$
through Lemma~\ref{lemma:level-crossing-prob}.
We first note that $\eta(\tilde{\ttt})$ is a Lipschitz continuous function,
\begin{equation}\label{eq:lipschiz-eta-stop}
	|
\eta(\tilde{\ttt})-\eta(\tilde{\ttt}')
	|\leq |l_{T\wedge \tau}(\tilde{\ttt}) - l_{T\wedge \tau}(\tilde{\ttt}')|
	\leq \tau \kappa_0 \norm{\tilde{\ttt}-\tilde{\ttt}'}.
\end{equation}
We further obtain the marginal tail probability of $\eta(\tilde{\ttt})$ through the next lemma.
\begin{lemma}\label{lemma:likelihood-ratio-crossing}
	For all $\tilde{\ttt}\neq \ttt$, and all constant $A>0$, we have
	$$\PP_{\ttt}\left(
	l_{T\wedge \tau}(\tilde{\ttt}) - l_{T\wedge \tau}(\ttt)\geq A
	\right)\leq e^{-A}$$
\end{lemma}
We take $A=h(c)-1$ in the above lemma and obtain
\begin{equation}
	\PP_{\ttt}\left(
	\eta(\tilde{\ttt})\geq h(c)-1
	\right)\leq e^{-h(c)+1}
\end{equation}
Combining the above display with \eqref{eq:lipschiz-eta-stop} and Lemma~\ref{lemma:level-crossing-prob}, we arrive at
\begin{equation}\label{eq:tail-LRT}
	\PP_{\ttt}\left(
		\sup_{\tilde{\ttt}\in W_{ji} } \eta(\tilde{\ttt})> h(c)
		\right)\leq O(\tau^{K-1} e^{-h(c)}).
\end{equation}
We combine \eqref{eq:tail-LRT},\eqref{eq:two-terms-split} and Lemma~\ref{lemma:t-bound} and arrive at
\begin{equation}
\begin{split}
			&\PP_{\ttt}\Big(\sup_{\tilde{\ttt}\in W_{ji}}l_{T_2}(\tilde{\ttt})-l_{T_2}(\ttt)> h(c)\Big)\\
		\leq &O(c^2)+ O(e^{-|\log c|-|\log c|^{1-\alpha}+(K-1)\log \tau})\\
= & O(c^2)+O(c e^{-|\log c|^{1-\alpha} +\tup (K-1) \log |\log c|})\\
		= & o(c ).
		\end{split}
\end{equation}
This completes our analysis for $T_2$.
We proceed to the analysis of the policy $\pi_1$ and the stopping time $T_1$.
According to the definition of $T_1$ in \eqref{eq:t-2}, we can see that upon stopping,
\begin{equation}
\begin{split}
	&\max_{(i,j):1\leq i<j\leq K} \exp\Big[\min\Big\{\sup_{\tilde{\ttt}\in W_{i,j}}l_{T_1}(\tilde{\ttt})-\sup_{\ttt\in W}l_{T_1}(\ttt), {\sup_{\tilde{\ttt}\in W_{j,i}}l_{T_1}(\tilde{\ttt})-\sup_{\ttt\in W}l_{T_1}(\ttt)} \Big\}\\
\leq &
\sum_{(i,j):1\leq i<j\leq K}
	\exp\Big[\min\Big\{\sup_{\tilde{\ttt}\in W_{i,j}}l_{T_1}(\tilde{\ttt})-\sup_{\ttt\in W}l_{T_1}(\ttt), {\sup_{\tilde{\ttt}\in W_{j,i}}l_{T_1}(\tilde{\ttt})-\sup_{\ttt\in W}l_{T_1}(\ttt)} \Big\}
	\Big]\\
	\leq & e^{-h(c)}.
\end{split}
\end{equation}
Taking logarithm and rearranging terms in the above display, we have
\begin{equation}\label{eq:t-1-upon-stopping}
	\min_{1\leq i<j\leq K}\Big[\sup_{\ttt\in W}
l_n(\ttt)-\min\Big\{\sup_{\tilde{\ttt}\in W_{i,j}}l_n(
	\tilde{\ttt}), \sup_{\tilde{\ttt}\in W_{j,i}}l_n(\tilde{\ttt}) \Big\}
	\Big]\geq {h(c)}.
\end{equation}
With \eqref{eq:t-1-upon-stopping} we can follow similar derivations as those for \eqref{eq:two-terms-split} and arrive at
\begin{equation}
\begin{split}
 & \EE L_K(\bar{D}_{T_1})\\
 \leq &
	\sup_{\ttt'\in W} \rho(\ttt) \text{meas}(W)\\
	&\times \frac{K(K-1)}{2} \sup_{\ttt\in W} \max_{(i,j):\ttt\notin W_{j,i}} \PP_{\ttt}\Big(\sup_{\tilde{\ttt}\in W_{ji}}l_{T_1}(\tilde{\ttt})-l_{T_1}(\ttt)> h(c)\Big).
\end{split}
\end{equation}
The rest of the proof is similar as that for the stopping time $T_2$. We omit the details.

\bigskip
\subsection{Proof of Theorem~\ref{thm:stopping-time}}\label{sec:proof-stopping}
Let $\delta$ be an arbitrary positive number, we can find an upper bound for the expectation of a stopping time $T$ as follows.
\begin{equation}\label{eq:exp-upper}
	\begin{split}
		&\EE T \\
		= & \sum_{m=0}^{\infty}\EE\left[T; m(1+\delta)t_c(\Theta)\leq T< (m+1)(1+\delta)t_c(\Theta) \right]\\
		\leq & (1+\delta)\EE t_c(\Theta) + \sum_{m=1}^{\infty }\EE\left[T; m(1+\delta)t_c(\Theta)\leq T< (m+1)(1+\delta)t_c(\Theta) \right]\\
		\leq &  (1+\delta)\EE t_c(\Theta) \\
		&+ (1+\delta)\max_{\ttt\in W} t_c(\ttt) \sum_{m=1}^{\infty }(m+1)\PP\left( m(1+\delta)t_c(\Theta)\leq T< (m+1)(1+\delta)t_c(\Theta) \right)\\
		\leq &(1+\delta)\EE t_c(\Theta) \\
		& +
		(1+\delta)\max_{\theta\in W} t_c(\ttt) \sum_{m=1}^{\infty }(m+1)\max_{\ttt\in W}\PP_{\ttt}\left( m(1+\delta)t_c(\ttt)\leq T< (m+1)(1+\delta)t_c(\ttt) \right)
	\end{split}
\end{equation}
We proceed to an upper bound for the probability in the above sum for $T=T_i$ ($i=1,2$). We start with $T=T_2$.
We split the probability for $m\geq 1$,
\begin{equation}\label{eq:another-two-terms}
\begin{split}
&\PP_{\ttt}\left( m(1+\delta)t_c(\ttt)\leq T_2 < (m+1)(1+\delta)t_c(\ttt) \right)\\
\leq &\PP_{\ttt}\Big( m(1+\delta)t_c(\ttt)\leq T_2 < (m+1)(1+\delta)t_c(\ttt),\\
&\qquad \max_{ m(1+\delta)\delta_2 t_{c}(\ttt)\leq t\leq m(1+\delta)t_c(\ttt) }\norm{\hat{\ttt}^{(t)}-\ttt}\leq |\log c|^{-\delta_1} \Big)\\
&+\PP_{\ttt}\Big(\max_{m(1+\delta)\delta_2 t_{c}(\ttt)\leq t\leq m(1+\delta)t_c(\ttt) }
 \norm{\hat{\ttt}^{(t)}-\ttt}\geq |\log c|^{-\delta_1}
\Big),
\end{split}	
\end{equation}
where we choose $\delta_1=\frac{\delta_0}{8}$ and $\delta_2=|\log c|^{-\delta_0/2}$, and $\delta_0$ is defined in Assumption~\ref{assump:randomized}. The second term on the above display is bounded from above according to Lemma~\ref{lemma:mle}, where we set
$n:=m(1+\delta)\delta_2 t_{c}(\ttt),
$ $m:=m(1+\delta)t_c(\ttt)$, $\varepsilon_{\lambda}=\Omega(|\log c|^{-\frac{1}{2}+\delta_0})$ and $\varepsilon_1 = |\log c|^{-\delta_1}$, and arrive at
\begin{equation}\label{eq:first-term-other}
\begin{split}
	&	\PP_{\ttt}\Big(\max_{m(1+\delta)\delta_2 t_{c}(\ttt)\leq t\leq m(1+\delta)t_c(\ttt) }
 \norm{\hat{\ttt}^{(t)}-\ttt}\geq |\log c|^{-\delta_1}
\Big)\\
\leq & e^{-\Omega(m(1+\delta)\delta_2 t_{c}(\ttt)|\log c|^{-4\delta_1}|\log c|^{-1+2\delta_0} ) }\times O(m^{K-1} |\log c|^{K-1})\\
=& e^{-\Omega( m |\log c|^{2\delta_0-4\delta_1}\delta_2 ) } O(m^{K-1}|\log c|^{K-1})\\
=& e^{-\Omega( m |\log c|^{\delta_0})  } O(m^{K-1}|\log c|^{K-1}).
\end{split}
\end{equation}
We proceed to the first term on the right-hand side of \eqref{eq:another-two-terms}.
For $m\geq 1$, we can see that $T_2> m(1+\delta)t_c(\ttt)$
implies that there exists $(i,j)$ such that $|\sup_{\tilde{\ttt}\in W_{i,j}}l_n(\tilde{\ttt})-\sup_{\ttt'\in W_{j,i}}l_n(\ttt')|\leq h(c)$ for $n=(1+\delta)m t_c(\ttt)$. Without loss of generality, we assume that $\ttt\in W_{i,j}$, then $T_2> m(1+\delta)t_c(\ttt)$  further implies
$l_n({\ttt})-\sup_{\ttt'\in W_{j,i}}l_n(\ttt')\leq h(c)$.
Therefore, an upper bound for the first term on the right-hand side of \eqref{eq:another-two-terms} is
\begin{equation}\label{eq:first-term-t2}
\begin{split}
		&\PP_{\ttt}\Big( m(1+\delta)t_c(\ttt)\leq T_2\leq (m+1)(1+\delta)t_c(\ttt);\\
		&\qquad\qquad  \max_{m(1+\delta)\delta_2 t_{c}(\ttt)\leq t\leq m(1+\delta)t_c(\ttt) }
		\norm{\hat{\ttt}^{(n)}-\ttt}\leq |\log c|^{-\delta_1}\Big)\\
		\leq &
		\PP_{\ttt}\Big(l_n({\ttt})-\sup_{\ttt'\in W_{j,i}}l_n(\ttt')\leq h(c);
		\max_{m(1+\delta)\delta_2 t_{c}(\ttt)\leq t\leq m(1+\delta)t_c(\ttt) }
		\norm{\hat{\ttt}^{(n)}-\ttt}\leq |\log c|^{-\delta_1}\Big),	
\end{split}
\end{equation}
We present an upper bound for the above display in the next lemma.
\begin{lemma}\label{lemma:sample-size-prob-bound}
If the strategy $\lambda^*(\hat{\ttt}^{(t)})$ is adopted with probability $1-o(1)$ uniformly for $mt_{c}(\ttt)(1+\delta)\delta_2 \leq t\leq  m(1+\delta)t_c(\ttt)$.
Then
\begin{equation}
	\begin{split}
		&\PP_{\ttt}\left(l_n({\ttt})-\sup_{\ttt'\in W_{j,i}}l_n(\ttt')\leq h(c);
		\max_{m(1+\delta)\delta_2 t_{c}(\ttt)\leq t\leq m(1+\delta)t_c(\ttt) }
		\norm{\hat{\ttt}^{(n)}-\ttt}\leq |\log c|^{-\delta_1}\right)\\
		\leq & e^{-\Omega(m|\log c|)}\times O(|\log c|^{K-1}m^{K-1}), 	
	\end{split}
\end{equation}
where $n=(1+\delta)m t_c(\ttt)$.
\end{lemma}
We combine the above lemma with \eqref{eq:first-term-other} and \eqref{eq:another-two-terms}, we arrive at
\begin{equation}
\begin{split}
		&\PP_{\ttt}\left( m(1+\delta)t_c(\ttt)\leq T_2< (m+1)(1+\delta)t_c(\ttt) \right)\\
	\leq &(e^{-\Omega(m |\log c|)} + e^{-\Omega( m |\log c|^{\delta_0})  }) \times O(m^{K-1}|\log c|^{K-1}).
\end{split}
\end{equation}
This, together with \eqref{eq:exp-upper} gives
\begin{equation}
\begin{split}
		& \EE T_2\\
	\leq & (1+\delta)\EE t_c(\Theta) \\
	&+ O(|\log c|)\times \sum_{m=1}^{\infty} (m+1) \{(e^{-\Omega(m |\log c|)} + e^{-\Omega( m |\log c|^{\delta_0})  }) \times O(m^{K-1}|\log c|^{K-1}) \}]\\
	\leq & (1+\delta)\EE t_c(\Theta) + o(|\log c|).
\end{split}
\end{equation}
This completes our analysis for $T_2$. We proceed to the analysis of $T_1$.
We can see that the event $T_1> n$ implies that
$$\sum_{(i,j)}\exp\Big[\min\Big\{\sup_{\tilde{\ttt}\in W_{i,j}}l_n(\tilde{\ttt})-\sup_{\ttt\in W}l_n(\ttt), {\sup_{\tilde{\ttt}\in W_{j,i}}l_n(\tilde{\ttt})-\sup_{\ttt\in W}l_n(\ttt)} \Big\}
	\Big]> e^{-h(c)},$$
which further implies that
\begin{equation*}
		K(K-1)\max_{(i,j)}\exp\Big[\min\Big\{\sup_{\tilde{\ttt}\in W_{i,j}}l_n(\tilde{\ttt})-\sup_{\ttt\in W}l_n(\ttt), {\sup_{\tilde{\ttt}\in W_{j,i}}l_n(\tilde{\ttt})-\sup_{\ttt\in W}l_n(\ttt)} \Big\}
	\Big]> e^{-h(c)}.
\end{equation*}

Simplifying the above display, we can see it is equivalent to that there exists $(i,j)$ such that
$$
|\sup_{\tilde{\ttt}\in W_{i,j}}l_n(\tilde{\ttt})-\sup_{\ttt\in W_{j,i}}l_n(\ttt)| \leq h(c)+\log K(K-1).
$$
The analysis is similar for the stopping time $T_1$ to that of $T_2$ by replacing $h(c)$ by $h(c)+\log K(K-1)$ in the derivation following \eqref{eq:first-term-t2}. We omit the details.

\section*{Acknowledgement}

The authors would like to thank Dr. Jingchen Liu and  Dr. Zhiliang Ying for the helpful discussions. Xi Chen acknowledges the support from Adobe Data Science Research Award and Xiaoou Li acknowledges the support from  National Science Foundation (NSF) under the grant DMS-1712657.

\bibliographystyle{plainnat}
\bibliography{reference}
\newpage

\appendix
 \begin{center}
    {\LARGE\bf Supplement to ``Asymptotically Optimal Sequential Design for Rank Aggregation"}
\end{center}
  \medskip

In this supplement, we provide the proofs of all the lemmas in the main paper.
\section{Proof of Lemma~\ref{lemma:mle}}\label{sec:proof-lemma}
\begin{proof}
	We first note that
	\begin{equation}\label{eq:eq1}
		\PP_{\ttt}\Big(
		\sup_{n\leq t\leq m}\norm{\hat{\ttt}^{(t)}-\ttt}\geq \varepsilon_1
		\Big)
		\leq
		(m-n)\times
		\max_{n\leq t\leq m}
				\PP_{\ttt}
\Big(
		\norm{\hat{\ttt}^{(t)}-\ttt}\geq \varepsilon_1
		\Big).
	\end{equation}
Note that $\norm{\hat{\ttt}^{(t)}-\ttt}\geq \varepsilon_1$ implies that the maximized logliklihood outside $B(\ttt,\varepsilon_1)$ is greater than that inside $B(\ttt,\varepsilon_1)$. Therefore, we have
\begin{equation}
	\PP_{\ttt}
\Big(
		\norm{\hat{\ttt}^{(t)}-\ttt}\geq \varepsilon_1
		\Big)
		\leq \PP_{\ttt}
\Big(\sup_{\ttt'\in W\backslash B(\ttt,\varepsilon_1)}l_t(\ttt')-l_t(\ttt)\geq 0
		\Big).
\end{equation}
From \eqref{eq:eq1} and the above display, we can see that it is sufficient to show that
\begin{equation}\label{eq:suff-to-show}
	\PP_{\ttt}
\Big(\sup_{\ttt'\in W\backslash B(\ttt,\varepsilon_1)}l_t(\ttt')-l_t(\ttt)\geq 0
		\Big) = e^{-\Omega(n \epsilon_{\lambda}^2\varepsilon_1^4)}\times O(m^{K-1}).
\end{equation}
For each $\ttt'\in W\backslash B(\ttt,\varepsilon_1)$, we consider the martingale
\begin{equation}
	M_t(\ttt')=l_t(\ttt')- l_t(\ttt)+\sum_{l=1}^t\sum_{(i,j)} \lambda_l^{(i,j)} D^{i,j} (\ttt\|\ttt').
\end{equation}
Then,
\begin{equation}
	\PP_{\ttt}
\Big(l_t(\ttt')-l_t(\ttt)\geq -1
		\Big)
= \PP_{\ttt}
\Big(M_t(\ttt')\geq \sum_{l=1}^t\sum_{(i,j)} \lambda_l^{(i,j)} D^{i,j} (\ttt\|\ttt')-1
		\Big)
		.
\end{equation}
Note that for $\ttt\in  W\backslash B(\ttt,\varepsilon_1) $ and $\min_{l,(i,j)} \lambda_l^{(i,j)}\geq \varepsilon_{\lambda}$. Combine this with Assumption~\ref{assump:identifiability} we have
\begin{equation}
	\sum_{l=1}^t\sum_{(i,j)} \lambda_l^{(i,j)} D^{i,j} (\ttt\|\ttt')-1
	=\Omega(t\varepsilon_{\lambda} \varepsilon_1^2) - 1= \Omega(t\varepsilon_{\lambda} \varepsilon_1^2).
\end{equation}
Therefore,
\begin{equation}
	\PP_{\ttt}
\Big(l_t(\ttt')-l_t(\ttt)\geq -1
		\Big)
\leq  \PP_{\ttt}
\Big(M_t(\ttt')\geq \Omega(t\varepsilon_{\lambda} \varepsilon_1^2)
		\Big)
		.
\end{equation}
We apply Lemma~\ref{lemma:hoeffding} to the above display and arrive at
\begin{equation}\label{eq:l-marginal}
\begin{split}
	&\PP_{\ttt}
\Big(l_t(\ttt')-l_t(\ttt)\geq -1
		\Big)\\
		 \leq& \PP_{\ttt}
\Big(M_t(\ttt')
\geq  \Omega(t\varepsilon_{\lambda} \varepsilon_1^2)
		\Big)\\
\leq & e^{-\Omega\left(
		\frac{ (t\varepsilon_{\lambda} \varepsilon_1^2)^2 }{t}
		 \right)}\\
		  = & e^{-\Omega\left(
		 t \varepsilon_{\lambda}^2 \varepsilon_1^4
		 \right)}
\end{split}
\end{equation}
On the other hand, it is easy to see that $l_t(\ttt')-l_t(\ttt)$ is Lipschitz in $\ttt'$
\begin{equation}
|l_t(\ttt')-l_t(\ttt) - l_t(\tilde{\ttt})-l_t(\ttt)|\leq t\kappa_0 \norm{\ttt'-\tilde{\ttt} }.	
\end{equation}
Combining the above display with \eqref{eq:l-marginal}, and Lemma~\ref{lemma:level-crossing-prob}, we arrive at
\begin{equation}
	\PP_{\ttt}
\Big(\sup_{\ttt'\in W\backslash B(\ttt,\varepsilon_1)}l_t(\ttt')-l_t(\ttt)\geq 0
		\Big) = O(t^{K-1})\times e^{-\Omega\left(
		 t \varepsilon_{\lambda}^2 \varepsilon_1^4
		 \right)} \leq O(m^{K-1})\times e^{-\Omega\left(
		 n \varepsilon_{\lambda}^2 \varepsilon_1^4
		 \right)}.
\end{equation}
The above display implies \eqref{eq:suff-to-show}, which completes our proof.
\end{proof}
\bigskip

\section{Proof of Lemma~\ref{lemma:second-term}}
\begin{proof}
Note that %
\begin{equation}\label{eq:first-term-exp2}
	\PP(\Theta\in U_r, B^c)=\PP(\Theta\in U_r)\PP(B^c|\Theta\in U_r).
\end{equation}
We focus on the conditional probability
\begin{align}
	&\PP(B^c|\Theta\in U_r)\\
	&=\PP\Big(\frac{
\PP(\Theta\in U_r|\FF_T)
}{\max_{(i,j):W_{i,j}\cap U_r=\emptyset}\PP(\Theta\in W_{i,j}|\FF_T)}\leq\frac{c^{\frac{\delta}{10}}}{\varepsilon}
|\Theta\in U_r\Big)\\
&=\PP\Big(\exists (i,j) \mbox{ such that } W_{i,j}\cap U_r=\emptyset \text{ and } \frac{
\PP(\Theta\in U_r|\FF_T)
}{\PP(\Theta\in W_{i,j}|\FF_T)}\leq\frac{c^{\frac{\delta}{10}}}{\varepsilon}
|\Theta\in U_r\Big)\\
&\leq \sum_{(i,j):W_{i,j}\cap U_r=\emptyset}\PP\Big( \frac{
\PP(\Theta\in U_r|\FF_T)
}{\PP(\Theta\in W_{i,j}|\FF_T)}\leq\frac{c^{\frac{\delta}{10}}}{\varepsilon}
|\Theta\in U_r\Big).\label{eq:sum-small}
\end{align}
We proceed to find an upper bound for each term in the above sum. For each $(i,j)$ such that $W_{i,j}\cap U_r=\emptyset$, we split the probability
\begin{align}
	&\PP\Big( \frac{
\PP(\Theta\in U_r|\FF_T)
}{\PP(\Theta\in W_{i,j}|\FF_T)}\leq\frac{c^{\frac{\delta}{10}}}{\varepsilon}
|\Theta\in U_r\Big)\\
&= \PP\Big( \frac{
\PP(\Theta\in U_r|\FF_T)
}{\PP(\Theta\in W_{i,j}|\FF_T)}\leq\frac{c^{\frac{\delta}{10}}}{\varepsilon}
, R_{i,j}=0|\Theta\in U_r\Big)\\
&+  \PP\Big( \frac{
\PP(\Theta\in U_r|\FF_T)
}{\PP(\Theta\in W_{i,j}|\FF_T)}\leq\frac{c^{\frac{\delta}{10}}}{\varepsilon}
, R_{i,j}=1|\Theta\in U_r\Big).\label{eq:sum-d}
\end{align}
Note that $W_{i,j}\cap U_r=\emptyset$ implies $\theta_i<\theta_j$ for all $\ttt\in U_r$. Consequently,
\begin{equation}\label{eq:same-d-ub}
\begin{split}
	&\PP\Big( \frac{
\PP(\Theta\in U_r|\FF_T)
}{\PP(\Theta\in W_{i,j}|\FF_T)}\leq\frac{c^{\frac{\delta}{10}}}{\varepsilon}
, R_{i,j}=1|\Theta\in U_r\Big)\\
 \leq& \PP\Big( R_{i,j}
=1|\Theta\in U_r\Big) =\E\Big[
I(\Theta_{i}<\Theta_{j}) R_{i,j}|\Theta\in U_r
\Big].
\end{split}
\end{equation}
Plug the above upper bound into \eqref{eq:sum-d}, we have
\begin{align*}
	&\PP\Big( \frac{
\PP(\Theta\in U_r|\FF_T)
}{\PP(\Theta\in W_{i,j}|\FF_T)}\leq\frac{c^{\frac{\delta}{10}}}{\varepsilon}
|\Theta\in U_r\Big)\\
&\leq \E\Big[
I(\Theta_{i}<\Theta_{j}) R_{i,j}|\Theta\in U_r
\Big] + \PP\Big( \frac{
\PP(\Theta\in U_r|\FF_T)
}{\PP(\Theta\in W_{i,j}|\FF_T)}\leq\frac{c^{\frac{\delta}{10}}}{\varepsilon}
, R_{i,j}=0|\Theta\in U_r\Big).
\end{align*}
We further plug the above display into \eqref{eq:sum-small} and get
\begin{align}
	&\PP\Big( \frac{
\PP(\Theta\in U_r|\FF_T)
}{\PP(\Theta\in W_{i,j}|\FF_T)}\leq\frac{c^{\frac{\delta}{10}}}{\varepsilon}
|\Theta\in U_r\Big)\\
\leq& \sum_{i\neq j}\E\Big[
I(\Theta_{i}<\Theta_{j}) R_{i,j}|\Theta\in U_r
\Big] \\
&+ \sum_{W_{i,j}\cap U_r=\emptyset}\PP\Big( \frac{
\PP(\Theta\in U_r|\FF_T)
}{\PP(\Theta\in W_{i,j}|\FF_T)}\leq\frac{c^{\frac{\delta}{10}}}{\varepsilon}
, R_{i,j}=0|\Theta\in U_r\Big).
\label{eq:lb-first-simp}
\end{align}
Recall the definition of $L_K=\sum_{(i,j)}I(\Theta_i<\Theta_j)R_{i,j}$, we find that the first term on the right-hand side of the above inequality is
\begin{equation}
	\sum_{i\neq j}\E\Big[
I(\Theta_{i}<\Theta_{j}) R_{i,j}|\Theta\in U_r
\Big] =\E[L_K|\Theta\in U_r].
\end{equation}
Consequently, \eqref{eq:lb-first-simp} can be further simplified as
\begin{align}
	&\PP\Big( \frac{
\PP(\Theta\in U_r|\FF_T)
}{\PP(\Theta\in W_{i,j}|\FF_T)}\leq\frac{c^{\frac{\delta}{10}}}{\varepsilon}
|\Theta\in U_r\Big)\\
&\leq\E[L_K|\Theta\in U_r] + \sum_{W_{i,j}\cap U_r=\emptyset}\PP\Big( \frac{
\PP(\Theta\in U_r|\FF_T)
}{\PP(\Theta\in W_{i,j}|\FF_T)}\leq\frac{c^{\frac{\delta}{10}}}{\varepsilon}
, R_{i,j}=0|\Theta\in U_r\Big).\label{eq:lb-first-simp}
\end{align}
We proceed to an upper bound of the second term on the right-hand side of the above inequality.
For each $(i,j)$ such that $W_{i,j}\cap U_r=\emptyset$, we consider the following two probability measures for $t=1,2,...$
\begin{align*}
	\tilde{\PP}(X_{1:t},a_{1:t}\in \cdot)&\triangleq\PP(
	X_{1:t},a_{1:t}\in \cdot|\Theta\in U_r
	)=\frac{\PP(X_{1:t},a_{1:t}\in \cdot\cap \Theta\in U_r)}{{\PP}(\Theta\in U_r)  }\\
	\QQ_{i,j}(X_{1:t},a_{1:t}\in \cdot)&\triangleq\PP(
	X_{1:t},a_{1:t}\in \cdot|\Theta\in W_{i,j}
	)=\frac{\PP(X_{1:t},a_{1:t}\in \cdot\cap \Theta\in W_{i,j})}{{\PP}(\Theta\in W_{i,j})  },
\end{align*}
where we write $X_{1:t}$ and $a_{1:t}$ for $X_1,....,X_t$ and $a_{1},...,a_t$.
Then, the Radon-Nikodym derivative upon stopping is
\begin{equation}\label{eq:RN}
	\frac{d\tilde{\PP}}{d\QQ_{i,j}}
	=
	\frac{\PP(\Theta\in U_r|\FF_{T} )/ \PP(\Theta\in U_r)
	}{\PP(\Theta\in W_{i,j}|\FF_{T})/ \PP(\Theta\in W_{i,j})}.
\end{equation}
We have
\begin{align}
	&\PP\Big( \frac{
\PP(\Theta\in U_r|\FF_T)
}{\PP(\Theta\in W_{i,j}|\FF_T)}\leq\frac{c^{\frac{\delta}{10}}}{\varepsilon}
, R_{i,j}=0|\Theta\in U_r\Big)\\
=&\tilde{\PP}\left(\frac{
\PP(\Theta\in U_r|\FF_T)
}{\PP(\Theta\in W_{i,j}|\FF_T)}\leq\frac{c^{\frac{\delta}{10}}}{\varepsilon}
, R_{i,j}=0\right)\\
=& \E^{\QQ_{i,j}}
\Big[
\frac{d\tilde{\PP}}{d\QQ_{i,j}}; \frac{\PP(\Theta\in U_r|\FF_T)
}{\PP(\Theta\in W_{i,j}|\FF_T)}\leq\frac{c^{\frac{\delta}{10}}}{\varepsilon}
, R_{i,j}=0
\Big].\label{eq:def-QQ}
\end{align}
We plug \eqref{eq:RN} into the above display
\begin{align*}
	&\PP\Big( \frac{
\PP(\Theta\in U_r|\FF_T)
}{\PP(\Theta\in W_{i,j}|\FF_T)}\leq\frac{c^{\frac{\delta}{10}}}{\varepsilon}
, R_{i,j}=0|\Theta\in U_r\Big)\\
=& \E^{\QQ_{i,j}}
\Big[
\frac{\PP(\Theta\in U_r|\FF_{T} )/ \PP(\Theta\in U_r)
	}{\PP(\Theta\in W_{i,j}|\FF_{T})/ \PP(\Theta\in W_{i,j})}; \frac{\PP(\Theta\in U_r|\FF_T)
}{\PP(\Theta\in W_{i,j}|\FF_T)}\leq\frac{c^{\frac{\delta}{10}}}{\varepsilon}, R_{i,j}=0
\Big].
\end{align*}
The above expression is further bounded above by
\begin{equation}
	\PP\Big( \frac{
\PP(\Theta\in U_r|\FF_T)
}{\PP(\Theta\in W_{i,j}|\FF_T)}\leq\frac{c^{\frac{\delta}{10}}}{\varepsilon}
, R_{i,j}=0|\Theta\in U_r\Big)\leq \frac{\PP(\Theta\in W_{i,j})}{\PP(\Theta\in U_r)}\frac{c^{\frac{\delta}{10}}}{\varepsilon} \QQ_{i,j}(R_{i,j}=0).
\end{equation}
According to the definition of $\QQ_{i,j}$ in \eqref{eq:def-QQ}, the above display implies
\begin{equation}
\begin{split}
	&\PP\Big( \frac{
\PP(\Theta\in U_r|\FF_T)
}{\PP(\Theta\in W_{i,j}|\FF_T)}\leq\frac{c^{\frac{\delta}{10}}}{\varepsilon}
, R_{i,j}=0|\Theta\in U_r\Big)\\
\leq & \frac{\PP(\Theta\in W_{i,j})}{\PP(\Theta\in U_r)}\frac{c^{\frac{\delta}{10}}}{\varepsilon} \PP(R_{i,j}=0|\Theta\in W_{i,j} ).
\end{split}
\end{equation}
Because $\theta_i>\theta_j$ for all $\ttt\in W_{i,j}$, we  have
\begin{align*}
	&\PP\Big( \frac{
\PP(\Theta\in U_r|\FF_T)
}{\PP(\Theta\in W_{i,j}|\FF_T)}\leq \frac{c^{\frac{\delta}{10}}}{\varepsilon}
, R_{i,j}=0|\Theta\in U_r\Big)\\
&\leq \frac{c^{\frac{\delta}{10}}}{\varepsilon}\frac{\PP(\Theta\in W_{i,j})}{\PP(\Theta\in U_r)}\EE[I(\Theta_i>\Theta_j)(1-R_{i,j})|\Theta\in W_{i,j}] \\
& =\frac{c^{\frac{\delta}{10}}}{\varepsilon}\frac{1}{\PP(\Theta\in U_r)}\EE[I(\Theta_i>\Theta_j)(1-R_{i,j}); \Theta\in W_{i,j}]
\end{align*}
Now we plug the above inequality into \eqref{eq:lb-first-simp}. We have
\begin{align*}
	&\PP\Big( \frac{
\PP(\Theta\in U_r|\FF_T)
}{\PP(\Theta\in W_{i,j}|\FF_T)}\leq\frac{c^{\frac{\delta}{10}}}{\varepsilon}
|\Theta\in U_r\Big)\\
\leq&\sum_{i\neq j}\E\Big[
I(\Theta_{i}<\Theta_{j}) R_{i,j}|\Theta\in U_r
\Big] + \frac{c^{\frac{\delta}{10}}}{\varepsilon}\frac{1}{\PP(\Theta\in U_r)}\sum_{W_{i,j}\cap U_r=\emptyset}\EE[I(\Theta_i>\Theta_j)(1-R_{i,j})]\\
& \leq \frac{1+\frac{c^{\frac{\delta}{10}}}{\varepsilon}}{\PP(\Theta\in U_r)} \E L_K(R).
\end{align*}
Recall that the policy considered here is satisfies $ \E L_K(R)\leq \varepsilon$. Consequently,
\begin{equation}\label{eq:second-term-exp}
	\PP\Big( \frac{
\PP(\Theta\in U_r|\FF_T)
}{\PP(\Theta\in W_{i,j}|\FF_T)}\leq \frac{c^{\frac{\delta}{10}}}{\varepsilon}
|\Theta\in U_r\Big)\leq \frac{1+\frac{c^{\frac{\delta}{10}}}{\varepsilon}}{\PP(\Theta\in U_r)}\varepsilon
\end{equation}
Therefore, \eqref{eq:first-term-exp2} is bounded from above by
\begin{equation}\label{eq:second-term-ub}
\PP(\Theta\in U_r, B^c) \leq {(1+\frac{c^{\frac{\delta}{10}}}{\varepsilon})} \varepsilon.	
\end{equation}
\end{proof}
\bigskip

\section{Proof of Lemma~\ref{lemma:transform-first-term}}
\begin{proof}
We first find an upper bound of $$\frac{\PP(\Theta\in U_r|\FF_{t})
}{\max_{(i,j):W_{i,j}\cap U_r=\emptyset}\PP(\Theta\in W_{i,j}|\FF_{t})}.$$ For the denominator,  we have
\begin{align*}
\max_{(i,j):W_{i,j}\cap U_r=\emptyset}\PP(\Theta\in W_{i,j}|\FF_{t})\geq \PP(\Theta\in W_{i^*_t,j^*_t}|\FF_{t})\geq \PP(\Theta\in W_{i^*_t,j^*_t}\cap B(\ttt^*,s)|\FF_{t}).
\end{align*}
Therefore,
\begin{eqnarray}
&&\frac{\PP(\Theta\in U_r|\FF_{t})
}{\max_{W_{i,j}\cap U_r=\emptyset}\PP(\Theta\in W_{i,j}|\FF_{t})}\\
&\leq& \frac{\PP(\Theta\in U_r|\FF_{t})}{\PP(\Theta\in W_{i^*_t,j^*_t}\cap B(\ttt^*,s)|\FF_{t})}\\
&=& \frac{\int_{U_r}\prod_{i=1}^t f^{a_i}_{\ttt}(X_i)\rho(\ttt)d\ttt }{\int_{W_{i^*_t,j^*_t}\cap B(\ttt^*_t,s)}\prod_{i=1}^t f^{a_i}_{\ttt}(X_i)\rho(\ttt)d\ttt }.
\end{eqnarray}
The above display  is further bounded from above by
\begin{equation}\label{eq:ub-stat}
	\frac{\PP(\Theta\in U_r)\max_{\ttt\in U_r}\prod_{i=1}^t f^{a_i}_{\ttt}(X_i)}{\PP(\Theta\in W_{i^*_t,j^*_t}\cap B(\ttt^*,s))\min_{\ttt\in W_{i^*_t,j^*_t}\cap B(\ttt^*_t,s)}\prod_{i=1}^t f^{a_i}_{\ttt}(X_i)}.
\end{equation}
According to Assumption~\ref{assump:compact}, we have
\begin{equation}\label{eq:lb-prob}
	\PP(\Theta\in W_{i^*_t,j^*_t}\cap B(\ttt^*_t,s))\geq \delta_{\rho}m(W_{i^*,j^*}\cap B(\ttt^*_t,s)) \geq \delta_{\rho}\delta_b s^{K-1},
\end{equation}
and for any $\ttt'\in B(\ttt^*_t,s)$,
\begin{equation}\label{eq:score-bound}
	|l^{\vec{a}}_{t}(\ttt')-l^{\vec{a}}_{t}(\ttt^*_t)|\leq t\kappa_0 \Vert \ttt'-\ttt^*_t\Vert \leq t\kappa_0s.
\end{equation}
Combining \eqref{eq:lb-prob} and \eqref{eq:score-bound} and \eqref{eq:ub-stat}, we have
 \begin{equation}
 \log\Big(	\frac{\PP(\Theta\in U_r|\FF_{t})
}{\max_{W_{i,j}\cap U_r=\emptyset}\PP(\Theta\in W_{i,j}|\FF_{t})}\Big)
\leq
\max_{\ttt'\in U_{r}} l^{\vec{a}}_t(\ttt') - l^{\vec{a}}_t(\ttt^*_t) + t\kappa_0 s - \log (\delta_{\rho}\delta_b s^{K-1}).
 \end{equation}
Therefore, we arrive at an upper bound for \eqref{eq:first-term}.
\begin{align}
	&\PP_{\ttt}\Big(
\max_{1\leq t\leq t(\ttt)}\frac{\PP(\Theta\in U_r|\FF_{t})
}{\max_{W_{i,j}\cap U_r=\emptyset}\PP(\Theta\in W_{i,j}|\FF_{t})}> \frac{c^{\frac{\delta}{10}}}{\varepsilon}\Big)\\
&\leq \PP_{\ttt}\Big(
\max_{\max_{1\leq t\leq t(\ttt), \ttt'\in U_{r}}} l^{\vec{a}}_t(\ttt') - l^{\vec{a}}_t(\ttt^*_t) + t\kappa_0 s - \log (\delta_{\rho}\delta_b s^K)>\log( \frac{c^{\frac{\delta}{10}}}{\varepsilon})
\Big)\\
&= \PP_{\ttt}\Big(
\max_{\ttt'\in U_{r}}\max_{\max_{1\leq t\leq t(\ttt)}} \{l^{\vec{a}}_t(\ttt') - l^{\vec{a}}_t(\ttt^*_t) + t\kappa_0 s \}>\log( \frac{c^{\frac{\delta}{10}}}{\varepsilon})+\log (\delta_{\rho}\delta_b s^{K-1})\Big)\label{eq:ub-max-l}
\end{align}
Recall the definition of $M_t(\ttt')$ in \eqref{eq:mart}. We can see that $l_t^{\vec{a}}(\ttt')-l_t^{\vec{a}}(\ttt^*_t)+t\kappa_0 s\geq \log( \frac{c^{\frac{\delta}{10}}}{\varepsilon})+\log (\delta_{\rho}\delta_b s^{K-1})$ is equivalent to
\begin{equation}
\begin{split}
	& M_t(\ttt')\\
\geq & \log( \frac{c^{\frac{\delta}{10}}}{\varepsilon})+\log (\delta_{\rho}\delta_b s^{K-1})-\sum_{n=1}^t\sum_{(i,j)}\lambda_n^{i,j}D^{i,j}(\ttt\|\ttt^*_t)\\
& + \sum_{n=1}^t\sum_{(i,j)}\lambda_n^{i,j}D^{i,j}(\ttt\|\ttt')-t\kappa_0 s,
\end{split}
\end{equation}
which further implies
\begin{equation}\label{eq:implies1}
	M_t(\ttt')\geq \log( \frac{c^{\frac{\delta}{10}}}{\varepsilon})+\log (\delta_{\rho}\delta_b s^{K-1})-\sum_{n=1}^t\sum_{(i,j)}\lambda_n^{i,j}D^{i,j}(\ttt\|\ttt^*_t) -t\kappa_0 s.
\end{equation}
\begin{lemma}\label{lemma:supp-KL}
For $t\geq 1$, we have
$$
\sum_{n=1}^t\sum_{(i,j)}\lambda_n^{i,j}D^{i,j}(\ttt\|\ttt^*_t)\leq t D(\ttt).
$$
\end{lemma}
With the aid of the above lemma, we see that \eqref{eq:implies1} implies that for $1\leq t\leq t_{c,\delta}(\ttt)$
\begin{equation}\label{eq:implies2}
	M_t(\ttt')\geq \log( \frac{c^{\frac{\delta}{10}}}{\varepsilon})+\log (\delta_{\rho}\delta_b s^{K-1})-t_{c,\delta}(\ttt)D(\ttt) -t_{c,\delta}(\ttt)\kappa_0 s.
\end{equation}
With our choice of $t_{c,\delta}(\ttt)$ and $D(\ttt)$, we have $t_{c,\delta}(\ttt)D(\ttt) = (1-2\delta/3) |\log c|$. Therefore, \eqref{eq:implies2} implies
\begin{equation}
		M_t(\ttt')\geq \log( \frac{c^{\frac{\delta}{10}}}{\varepsilon})+\log (\delta_{\rho}\delta_b s^{K-1})-(1-2\delta/3) |\log c|-t_{c,\delta}(\ttt)\kappa_0 s.
\end{equation}
As a consequence, \eqref{eq:ub-max-l} gives
\begin{align}
	&\PP_{\ttt}\Big(
\max_{1\leq t\leq t(\ttt)}\frac{\PP(\Theta\in U_r|\FF_{t})
}{\max_{W_{i,j}\cap U_r=\emptyset}\PP(\Theta\in W_{i,j}|\FF_{t})}>\frac{c^{\frac{\delta}{10}}}{\varepsilon}\Big)\\
\leq &
\PP_{\ttt}\Big(
\max_{1\leq t\leq t(\ttt),\ttt\in U_r}M_t(\ttt')\geq \log( \frac{c^{\frac{\delta}{10}}}{\varepsilon})+\log (\delta_{\rho}\delta_b s^{K-1})\\
&-(1-2\delta/3) |\log c|-t_{c,\delta}(\ttt)\kappa_0 s
\Big)\label{eq:ineq-m}
\end{align}
Now, we choose $s=|\log c|^{-1}$ and recall that $\varepsilon = c|\log c|^2$. We have
\begin{equation}
\begin{split}
& \log( \frac{c^{\frac{\delta}{10}}}{\varepsilon})+\log (\delta_{\rho}\delta_b s^{K-1})-(1-2\delta/3) |\log c|-t_{c,\delta}(\ttt)\kappa_0 s\\
= & \frac{\delta}{10}\log c - 2\log|\log c| +|\log c| - K\log |\log c| - (1-2\delta/3)|\log c|	+ O(1)\\
=& \frac{17\delta}{30} |\log c| -(K+2)\log |\log c|+O(1)\\
\geq &\frac{\delta}{2}|\log c|
\end{split}
\end{equation}
for $c$ sufficiently small. Combining this with \eqref{eq:ineq-m}, we have
\begin{equation*}
\begin{split}
	& \PP_{\ttt}\Big(
\max_{1\leq t\leq t(\ttt)}\frac{\PP(\Theta\in U_r|\FF_{t})
}{\max_{W_{i,j}\cap U_r=\emptyset}\PP(\Theta\in W_{i,j}|\FF_{t})}>\frac{c^{\frac{\delta}{10}}}{\varepsilon}\Big)\\
\leq &\PP_{\ttt}\Big(
\max_{1\leq t\leq t(\ttt),\ttt'\in U_r}M_t(\ttt')\geq \frac{\delta}{2}|\log c|
\Big).
\end{split}
\end{equation*}
\end{proof}
\bigskip

\section{Proof of Lemma~\ref{lemma:level-crossing-prob}}
\begin{proof}
We define a change of measure $Q$, under which the random field  $\{\zeta(\ttt):\ttt\in U\}$ is sampled as follows,
\begin{enumerate}
	\item Sample a random index $\upsilon\in U$ with a density function $h(\upsilon)\propto \PP(\zeta(\upsilon)>b-\gamma)$.
	\item Given the index $\upsilon$ generated in the first step, sample $\zeta(\upsilon)$ conditional on $\upsilon$ and $\zeta(\upsilon)>b-\gamma$.
	\item Sample $\{\zeta(\ttt):\ttt\neq \upsilon\}$ according to the conditional distribution $\{\zeta(\ttt):\ttt\neq \upsilon\}|\upsilon,\zeta(\upsilon)$ according to the original probability measure $\PP$.
\end{enumerate}
This change of measure admits the following Radon-Nikodym derivative
\begin{equation}\label{eq:radon}
	\frac{dQ}{d\PP}=\frac{m(A_{b-\gamma})}{\int_U \PP(\zeta(\ttt)>b-\gamma) d\ttt},
\end{equation}
where the set $A_{b-\gamma}=\{\ttt\in U: \zeta(\ttt)>b-\gamma \}$ is a random excursion set and $m(\cdot)$ denotes its Lebesgue measure. For a rigorous justification of \eqref{eq:radon} and some examples, see \cite{adler2012efficient,li2014generalized,li2015rare,li2016chernoff}
The  probability  of interest is
\begin{equation}\label{eq:prob-after-change}
\begin{aligned}
	&\PP\left(
\max_{\ttt\in U} \zeta(\ttt)\geq b \right)
= \EE^Q\left[
\frac{d\PP}{dQ}I(\sup_{\ttt\in U} \zeta(\ttt)>b)
\right]\\
= &\int_U \PP(\zeta(\ttt)>b-\gamma) d\ttt\times \EE^Q\left[
\frac{1}{m(A_{b-\gamma})}I(\sup_{\ttt\in U} \zeta(\ttt)>b)
\right]
\end{aligned}
\end{equation}
On the other hand, when the event $\sup_{\ttt\in U} \zeta(\ttt)>b$ happen, we let $\ttt^*=\arg\sup_{\ttt}\zeta(\ttt)$. Note that for $|\ttt-\ttt^*|\leq \frac{\gamma}{\kappa_L} $, $\zeta(\ttt)\geq \zeta(\ttt^*)-\kappa_L \frac{a}{\kappa_L}\geq b-\gamma.$ Therefore, the Lebesgue measure of $A_{b-\gamma}$ is bounded from below by $m(U\cap\{ \ttt:|\ttt-\ttt^*|\leq \frac{\gamma}{\kappa_L}\})\geq \frac{\gamma^{K-1} \delta_b}{\kappa_L^{K-1}}$.
Therefore,
$$ \EE^Q\left[
\frac{1}{m(A_{b-a})}I(\sup_{\ttt\in U} \zeta(\ttt)>b)
\right]\leq \frac{\kappa_L^{K-1}}{\gamma^{K-1} \delta_b}.$$ This, together with \eqref{eq:prob-after-change}, completes the proof.

\end{proof}
\bigskip

\section{Proof of Lemma~\ref{lemma:t-bound}}
\begin{proof}
According to the definition of $T_2$, we can see that the event $T_2> \tau$ implies that  there exists $(i,j)$ such that $|\sup_{\tilde{\ttt}\in W_{i,j}} l_{\tau}(\tilde{\ttt})- \sup_{\ttt' \in W_{j,i}}l_{\tau}(\ttt')|<h(c)$. Therefore,
\begin{equation}
	\PP_{\ttt}\left(
	T_2>\tau
	\right)
	\leq \PP_{\ttt}\left(
	\exists (i,j) \mbox{ such that } |\sup_{\tilde{\ttt}\in W_{i,j}} l_{\tau}(\tilde{\ttt})- \sup_{\ttt'\in W_{j,i}}l_{\tau}(\ttt')|<h(c)
	\right),
\end{equation}
which is further bounded from above by
\begin{equation}
	\PP_{\ttt}\left(
	T_2>\tau
	\right)
	\leq \sum_{(i,j)}\PP_{\ttt}\left(
 |\sup_{\tilde{\ttt}\in W_{i,j}} l_{\tau}(\tilde{\ttt})- \sup_{\ttt'\in W_{j,i}}l_{\tau}(\ttt')|<h(c)
	\right).
\end{equation}
For each $(i,j)$, we proceed to an upper bound of $\PP_{\ttt}\left(
 |\sup_{\tilde{\ttt}\in W_{ij}} l_{\tau}(\tilde{\ttt})- \sup_{\ttt'\in W_{ji}}l_{\tau}(\ttt')|<h(c)
	\right)$.
Without loss of generality, we assume that $\ttt\in W_{ij}$ and thus $\ttt\notin W_{ji}$. Then,
\begin{eqnarray}
	&\PP_{\ttt}\left(
 |\sup_{\tilde{\ttt}\in W_{ij}} l_{\tau}(\tilde{\ttt})- \sup_{\ttt'\in W_{ji}}l_{\tau}(\ttt')|<h(c)
	\right)\\
	\leq& \PP_{\ttt}\left(
 \sup_{\tilde{\ttt}\in W_{ij}} l_{\tau}(\tilde{\ttt})- \sup_{\ttt'\in W_{ji}}l_{\tau}(\ttt')<h(c)
	\right)\\
	\leq & \PP_{\ttt}\left(
  l_{\tau}({\ttt})- \sup_{\ttt'\in W_{ji}}l_{\tau}(\ttt')<h(c)
	\right)\\
	=&  \PP_{\ttt}\left(
 \sup_{\ttt' \in W_{ji}}l_{\tau}(\ttt')- l_{\tau}({\ttt})> -h(c)
	\right).
\end{eqnarray}
Therefore, it is sufficient to show that $\PP_{\ttt}\left(
 \sup_{W_{ji}}l_{\tau}(\ttt')- l_{\tau}({\ttt})> -h(c)
	\right) =O(c^2)$.
This is in the form of the level crossing probability. We will find an upper bound via Lemma~\ref{lemma:level-crossing-prob}.
We define the martingale,
\begin{equation}
	M_{t}(\ttt')= l_t(\ttt')-l_t(\ttt) +\sum_{n=1}^{t} \sum_{(i,j)} \lambda_n^{i,j} D^{i,j}(\ttt\|\ttt').
\end{equation}
Then
\begin{equation}
	\PP_{\ttt}\left(
 l_{\tau}(\ttt')- l_{\tau}({\ttt})> -h(c)-1
\right)
=
\PP_{\ttt}\left(
M_{\tau}(\ttt')\geq \sum_{n=1}^{\tau} \sum_{(i,j)} \lambda_n^{i,j} D(\ttt\|\ttt')-h(c)-1
\right).
\end{equation}
According to Assumption~\ref{assump:randomized}, we have
\begin{equation}
\begin{split}
	&\sum_{n=1}^{\tau}\sum_{(i,j)} \lambda_n^{i,j} D^{i,j}(\ttt\|\ttt')-h(c)\\
	\geq &\Omega(\tau |\log c|^{-\frac{1}{2}+\delta_0 }\max_{(i,j)}D^{i,j}(\ttt\|\ttt'))-h(c)\\
	=&\Omega(\tau |\log c|^{-\frac{1}{2}+\delta_0 })
\end{split}
\end{equation}
For the last equation in the above display, we used the fact that $h(c)= o(\tau |\log c|^{-\frac{1}{2}+\delta_0 })$ for our choice of $\tau$.
Consequently,
\begin{equation}
	\PP_{\ttt}\left(
 l_{\tau}(\ttt')- l_{\tau}({\ttt})> -h(c)-1
\right)
\leq
\PP_{\ttt}\left(
M_{\tau}(\ttt')\geq   \Omega(
	\tau |\log c|^{-\frac{1}{2}+\delta_0 }
	)
\right).
\end{equation}
Applying Lemma~\ref{lemma:hoeffding} to $M_{\tau}(\ttt')$, we have
\begin{equation}
\begin{split}
	& \PP_{\ttt}\left(
M_{\tau}(\ttt')\geq   \Omega(
	\tau |\log c|^{-\frac{1}{2}+\delta_0 }
	)
\right)\\
\leq & e^{
- \Omega(\frac{(\tau |\log c|^{-\frac{1}{2}+\delta_0 }
)^2}{\tau})}\\
= & e^{
- \Omega({\tau|\log c|^{-1+2\delta_0 }
)}}\\
= & e^{- \Omega(|\log c|^{2+2\delta_0 })}.
\end{split}
\end{equation}
Therefore,
\begin{equation}\label{eq:eta-marginal-tail}
	\PP_{\ttt}\left(
 l_{\tau}(\ttt')- l_{\tau}({\ttt})> -h(c)-1
\right)
\leq e^{- \Omega(|\log c|^{2+2\delta_0 })}.
\end{equation}
On the other hand, the random field $\eta(\ttt')= l_{\tau}(\ttt')- l_{\tau}({\ttt})$ is Lipschitz,
\begin{equation}\label{eq:eta-lip}
	|\eta(\ttt')- \eta(\tilde{\ttt})| = |l_{\tau}(\ttt')- l_{\tau}(\tilde{\ttt})|\leq \tau \kappa_0.
\end{equation}
We combine \eqref{eq:eta-marginal-tail}, \eqref{eq:eta-lip} and Lemma~\ref{lemma:level-crossing-prob} and arrive at
\begin{equation}
\begin{split}
	&\PP_{\ttt}\left(
 \sup_{W_{ji}}l_{\tau}(\ttt')- l_{\tau}({\ttt})> -h(c)
	\right)\\
	\leq & O(\tau^{K-1})\times  e^{- \Omega(|\log c|^{2+2\delta_0 })}\\
	= & O(e^{- \Omega(|\log c|^{2})})\\
	=& O(c^2).
\end{split}
\end{equation}
This completes our proof.
\end{proof}
\bigskip

\section{Proof of Lemma~\ref{lemma:likelihood-ratio-crossing}}
\begin{proof}
Consider the probability measure $\PP_{\tilde{\ttt}}$. The Radon-Nikodym derivative is
\begin{equation}
	\frac{d\PP_{\tilde{\ttt}}}{d\PP}=e^{l_{T\wedge \tau}(\tilde{\ttt})-l_{T\wedge \tau}({\ttt})}.
\end{equation}
Therefore,
\begin{equation}
\begin{split}
	&\PP\left(
	l_{T\wedge \tau}(\tilde{\ttt}) - l_{T\wedge \tau}(\ttt)\geq A
	\right)\\
	= &\EE_{\tilde{\ttt}}\left[
	e^{l_{T\wedge \tau}({\ttt})-l_{T\wedge \tau}(\tilde{\ttt})}; l_{T\wedge \tau}(\tilde{\ttt}) - l_{T\wedge \tau}(\ttt)\geq A
	\right]\\
	\leq & e^{-A}.
\end{split}
\end{equation}

\end{proof}

\bigskip

\section{Proof of Lemma~\ref{lemma:sample-size-prob-bound}}
\begin{proof}
	We first present a useful lemma, whose proof will be provided later in this section.
	\begin{lemma}\label{lemma:lip-kullback}
	There exists a positive constant $\kappa_D$ such that
	$$
	|D(\tilde{\ttt})-D(\ttt)|\leq \kappa_D \norm{\tilde{\ttt}-\ttt}
	$$
	for all $\tilde{\ttt},\ttt\in W$ such that $r(\ttt)=r(\tilde{\ttt})$.
	\end{lemma}
With the aid of the above lemma, and the assumption that the strategy $\lambda^*$ is adopted with probability $1-o(1)$, for $\max_{t}\norm{{\ttt}-\hat{\ttt}^{(t)}}\leq |\log c|^{-\delta_0/8}$ and $\ttt'\in W_{j,i}$, we have
\begin{equation}\label{eq:long-ineq}
\begin{split}
		& \sum_{t= (1+\delta)t_c(\ttt)m \delta_2}^{(1+\delta)t_c(\ttt)m}\sum_{(i,j)}\lambda_{t}^{(i,j)}D^{i,j}(\ttt\|{\ttt}')\\
		\geq &(1-o(1)) \sum_{t= (1+\delta)t_c(\ttt)m \delta_2}^{(1+\delta)t_c(\ttt)m}\sum_{(i,j)}\lambda_{t}^{*,(i,j)}D^{i,j}(\ttt\|{\ttt}')\\
		\geq &  (1-o(1))(1-|\log c|^{-\delta_1})\sum_{t= (1+\delta)t_c(\ttt)m \delta_2}^{(1+\delta)t_c(\ttt)m}\sum_{(i,j)}\lambda_{t}^{*,(i,j)}D^{i,j}(\hat{\ttt}_{t}\|{\ttt}')\\
		\geq & (1-o(1))(1-|\log c|^{-\delta_1})\sum_{t= (1+\delta)t_c(\ttt)m \delta_2}^{(1+\delta)t_c(\ttt)m}D(\hat{\ttt}_{t})\\
		\geq & (1-o(1))(1-|\log c|^{-\delta_1})[(1+\delta)t_c(\ttt)m-(1+\delta)t_c(\ttt)m \delta_2](D(\ttt)+O(|\log c|^{-\delta_1}))\\
	\geq &
	(1-o(1)) (1-|\log c|^{-\delta_1})(1-\delta_2)(1+\delta)t_c(\ttt)m D(\ttt)\\
	 =& (1-o(1))(1+\delta)t_c(\ttt)m D(\ttt).
\end{split}
\end{equation}
We explain the above derivation. The first inequality is due to the assumption that $\lambda^*$ is adopted with the probability $1-o(1)$. The second inequality is due to $\norm{\ttt -\hat{\ttt}^{(t)}}\leq |\log c|^{-\delta_1}$ and the Kullback-Leibler divergence $D^{i,j}(\ttt\|\ttt')$ is Lipschitz in $\ttt$. The third inequality is obtained according to the definition of the $D(\cdot)$ function. The fourth inequality is due to Lemma~\ref{lemma:lip-kullback}. The fifth and last inequalities are straightforward simplification of the previous lines.
Recall that $t_c(\ttt)D(\ttt)=|\log c|$ and $h(c)=|\log c|(1+|\log c|^{-\alpha})$, we can see $(1-o(1))(1+\delta)m |\log c|-h(c) = (1+o(1))\delta m|\log c|$. Therefore, we \eqref{eq:long-ineq} implies
\begin{equation}\label{eq:d-refined-bound}
	\sum_{t= (1+\delta)t_c(\ttt)m \delta_2}^{(1+\delta)t_c(\ttt)m}\sum_{(i,j)}\lambda_{t}^{(i,j)}D^{i,j}(\ttt\|{\ttt}')- h(c) \geq (1+o(1))\delta m|\log c|
\end{equation}
Now for each $\ttt'$ we define a martingale
$$
M_{n}(\ttt')=l_n(\ttt')-l_n(\ttt)+\sum_{t=1}^n \sum_{(i,j)}\lambda_t^{i,j}D^{i,j}(\ttt\|\ttt').
$$
Then, the probability of interest is
\begin{equation}
\begin{split}
	&\PP_{\ttt}\Big(
	\sup_{\ttt'\in W_{j,i}}\{ M_{n}(\ttt')- \sum_{t=1}^n \sum_{(i,j)}\lambda_t^{i,j}D^{i,j}(\ttt\|\ttt')\}\\
	&~~~~~~\geq - h(c);\max_{mt_{c}(\ttt)(1+\delta)\delta_2 \leq t\leq  m(1+\delta)t_c(\ttt)}
		\norm{\hat{\ttt}_n-\ttt}\leq |\log c|^{-\delta_1}
	\Big).
\end{split}
\end{equation}
According to \eqref{eq:d-refined-bound} and our choice of $n$, the above probability is bounded from above by
 $\PP_{\ttt}\left(
	\sup_{\ttt'\in W_{j,i}} M_{n}(\ttt') \geq (1+o(1))\delta|\log c|
	\right)$.
It is sufficient to show that
\begin{equation}
	\PP_{\ttt}\left(
	\sup_{\ttt'\in W_{j,i}} M_{n}(\ttt') \geq (1+o(1))\delta m |\log c|
	\right)\leq e^{-\Omega(m|\log c|)} O(|\log c|^{K-1}m^{K-1}).
\end{equation}
Recall that $n=m(1+\delta)t_c(\ttt)= O(m|\log c|)$. From Lemma~\ref{lemma:hoeffding}, we have that for each $\ttt'\in W_{j,i}$,
\begin{equation}\label{eq:marginal-m}
	\PP_{\ttt}\left(
	M_{n}(\ttt') \geq (1+o(1))\delta |\log c|-1
	\right)\leq e^{-\Omega( m |\log c|)}
\end{equation}
Also notice that $M_n(\ttt')$ is Lipshitz in $\ttt'$ with a Lipschitz constant of the order $O(n)$. With the aid of Lemma~\ref{lemma:level-crossing-prob}, and \eqref{eq:marginal-m} we have
\begin{equation}
\begin{aligned}
	&\PP_{\ttt}\left(
	M_{n}(\ttt') \geq  (1+o(1))\delta |\log c|
	\right)\leq e^{-\Omega(m|\log c|)} O(n^{K-1})\\
=& e^{-\Omega(m|\log c|)} O(|\log c|^{K-1}m^{K-1}).
\end{aligned}
\end{equation}
This completes our proof.

\end{proof}
\bigskip

\section{Proof of Lemma~\ref{lemma:supp-KL}}
\begin{proof}
According to the definition of $\ttt_t^*$ in \eqref{eq:thetastar}, we have
\begin{equation}
\begin{split}
&	\sum_{n=1}^t\sum_{(i,j)}\lambda_n^{i,j}D^{i,j}(\ttt\|\ttt^*_t)\\
= &\min_{\tilde{\ttt}:r(\tilde{\ttt})\neq r(\ttt)} \sum_{n=1}^t \sum_{(i,j)}\lambda_n^{i,j}D^{i,j}(\ttt\|\tilde{\ttt})\\
= & t \min_{\tilde{\ttt}:r(\tilde{\ttt})\neq r(\ttt)} \sum_{(i,j)}\frac{1}{t}\sum_{n=1}^t \lambda_n^{i,j}D^{i,j}(\ttt\|\tilde{\ttt})\\
\leq & t \max_{\lambda \in\triangle} \min_{\tilde{\ttt}:r(\tilde{\ttt})\neq r(\ttt)}  \sum_{(i,j)}\lambda^{i,j}D^{i,j}(\ttt\|\tilde{\ttt})\\
\end{split}
\end{equation}
The last inequality in the above display is due to
 $\frac{1}{t}\sum_{n=1}^t \lambda_n\in \triangle$.
We complete the proof by recalling the definition of $D(\ttt)=\max_{\lambda \in\triangle} \min_{\tilde{\ttt}:r(\tilde{\ttt})\neq r(\ttt)}  \sum_{(i,j)}\lambda^{i,j}D^{i,j}(\ttt\|\tilde{\ttt})$.
\end{proof}
\bigskip

\section{Proof of Lemma~\ref{lemma:lip-kullback}}
\begin{proof}
	Without loss of generality, we assume that $D(\tilde{\ttt})<D(\ttt)$. Then, $|D(\tilde{\ttt})-D(\ttt)| = D(\ttt)-D(\tilde{\ttt})$. Let $\lambda^*=\arg\max_{\lambda\in \triangle} \min_{r(\ttt')\neq r(\ttt) } \sum_{(i,j)} \lambda^{i,j} D^{i,j}(\ttt\|\ttt')$. Then, we have
	\begin{eqnarray*}
		&&D(\ttt)-D(\tilde{\ttt})\\
		&=&\min_{r(\ttt')\neq r({\ttt})} \sum_{(i,j)} \lambda^{*,i,j}D^{i,j}(\tilde{\ttt}\|\ttt')-\max_{\lambda\in\triangle}\min_{r(\ttt')\neq r(\tilde{\ttt})} \sum_{(i,j)} \lambda^{i,j}D^{i,j}(\tilde{\ttt}\|\ttt')  \\
		&\leq & \min_{r(\ttt')\neq r({\ttt})} \sum_{(i,j)} \lambda^{*,i,j}D^{i,j}({\ttt}\|\ttt')-\min_{r(\ttt')\neq r(\tilde{\ttt})} \sum_{(i,j)} \lambda^{*,i,j}D^{i,j}(\tilde{\ttt}\|\ttt')
	\end{eqnarray*}
	Note that for each $(i,j)$, $|D^{i,j}(\tilde{\ttt}\|\ttt') - D^{i,j}({\ttt}\|\ttt')| =O(\norm{\tilde{\ttt}-\ttt})$. Therefore, the above display further implies
	\begin{eqnarray*}
		&&D(\ttt)-D(\tilde{\ttt})\\
		&\leq & \min_{r(\ttt')\neq r({\ttt})} \sum_{(i,j)} \lambda^{*,i,j}D^{i,j}({\ttt}\|\ttt')-\min_{r(\ttt')\neq r({\ttt})} \sum_{(i,j)} \lambda^{*,i,j}D^{i,j}({\ttt}\|\ttt')  + O(\norm{\tilde{\ttt}-\ttt})\\
		&= & O(\norm{\tilde{\ttt}-\ttt}).
	\end{eqnarray*}
	This completes our proof.
\end{proof}

\end{document}